\documentclass[%
 reprint,
 amsmath,amssymb,
 aps,
pra,
floatfix,
]{revtex4-1}
\usepackage{graphicx,lipsum}% Include figure files
\usepackage{dcolumn}% Align table columns on decimal point
\usepackage{bm}% bold math
\usepackage[dvipsnames]{xcolor}
\usepackage{array,multirow}
\begin{document}

\preprint{APS/123-QED}
\title{Hierarchy of quantum correlations under non-Markovian dynamics}
%\title{Dynamics of quantum  correlations under the influence of  (non-)Markovian environmental interactions }% Force line breaks with \\

\author{K.G. Paulson\textsuperscript{a,b}}
\altaffiliation[paulsonkgeorg@gmail.com]{}%Lines break automatically or can be forced with \\
\author{Ekta Panwar\textsuperscript{c}}%
\email{panwar.1@iitj.ac.in}
\author{Subhashish Banerjee\textsuperscript{c}}
\email{subhashish@iitj.ac.in }
\author{R. Srikanth\textsuperscript{d}}
 \email{srik@poornaprajna.org }
\affiliation{Institute of Physics, Bhubaneswar-751005, India\textsuperscript{a}\\
Indian Institute of Science Education and Research Kolkata, Mohanpur-741246, West Bengal, India\textsuperscript{b}\\
Indian Institute of Technology Jodhpur, Jodhpur-342011, India\textsuperscript{c}\\
Poornaprajna Institute of Scientific Research, Sadashivnagar, Bengaluru-560080, India\textsuperscript{d}}
%\author{Ekta Panwar}
 %\email{}
%\affiliation{Indian Institute of Technology Jodhpur, Jodhpur 342011, India
%\textsuperscript{b,c}}
%\affiliation{
 %Third institution, the second for Charlie Author
%\affiliation{%


\date{\today}% It is always \today, today,

\begin{abstract}
We investigate the dynamics of quantum  correlations (QC) under the effects of reservoir memory,  as a resource for quantum information and computation tasks. Quantum correlations of two-qubit systems are used for implementing quantum teleportation successfully, and for investigating how  teleportation fidelity, violation of Bell-CHSH inequality, quantum steering  and entanglement are connected with each other under the influence of noisy environments. Both Markovian and non-Markovian channels are considered, and it is shown that the decay and revival of correlations follow the hierarchy of quantum correlations  in the state space. Noise tolerance of quantum correlations is checked for different types of unital and non-unital  quantum channels, with and without memory. The  quantum speed limit time $(\tau_{QSL})$ is investigated from the perspective of memory of quantum noise, and the corresponding dynamics is used to analyze the evolution of quantum correlations. We establish the connection between information backflow,  quantum speed limit time and dynamics of quantum correlations for non-Markovian quantum channels.
%\begin{description}
%\end{description}
\end{abstract}

\maketitle

%\tableofcontents

\section{\label{sec:level1} Introduction } 
Quantum correlations (QC) along with the superposition principle, triggered  the advances in the field  of quantum enabled science and technology. Right from its inception, quantum entanglement has exercised a pivotal role as a resource for manipulating the quantum information. In recent times, more general forms of non-classical correlations have been explored, and are widely used  as a resource for the successful implementation of various quantum information and computation protocols like, dense coding, teleportation,  key distribution, cryptography, parameter estimation, metrology ~\cite{ekert1991,bennett1992,bennett1993,goldenberg1995,suzuki2016,costa2016,cryptoswitch,PingPongM,tomogram}, both theoretically and experimentally. 
Entanglement is an inevitable resource~\cite{sbravi1,sbravi2} for achieving maximum possible success for a number of quantum information protocols (QIP), whereas  it is considered as a non-critical resource in the realization of some of the aforementioned protocols. Non-zero fidelity for quantum remote state preparation~\cite{dakic2012} can be achieved using separable states, wherein quantum discord enables the process. Recently, it has been shown that PPT bound entangled states are useful for quantum parameter estimation in noisy environment~\cite{toth2018}.

The study on multipartite quantum teleportation reveals that maximum degree of entanglement is not 
necessary~\cite{barasinski2019, paulson2019} to attain optimum teleportation fidelity. 
All these cases of QC as a resource for QIP~\cite{indranil, dhar2013, samya}  invite wide attention for the exploration of more general forms of non-classical correlations in bipartite and multipartite quantum systems, and their dynamics under decoherence.
Quantum decoherence is a phenomenon that occurs when quantum systems interact with their ambient environment, and is studied under the broader perspective of 
{\it {Open Quantum Systems}}~\cite{bp,sbbook}. Open quantum systems have been applied extensively in recent times to various facets of quantum information, condensed matter systems~\cite{weiss}, and relativistic as well as sub-atomic
physics
~\cite{unruh1,unruh2, richard,BK1,BK2,coherenceBK,sin2beta}. Coupling of the quantum system with the reservoir can be either weak or strong, leading to a wide range of dynamics. Quantum channels modeling these effects can be Markovian or non-Markovian, both unital as well as non-unital. 
The backflow of information from the reservoir to the system for a given non-Markovian quantum channel reveals many intriguing features of QC ~\cite{breuer2009measure,jyrki2,jyrki1}. In general, investigating the dynamics of quantum correlations of open quantum system, both Markovian as well as non-Markovian is pertinent, since the detailed study of coherence dynamics and correlations  of quantum states is  essential for the successful  implementation of quantum information and computation protocols~ \cite{jyrkinature,NMCrypto,Pradeep,QuThermo,NMDephasing1,JavidSupriyo,JavidCoh}. \newline
The hierarchy of quantum correlations of states of composite systems is known; to begin with, the classifications of quantum correlations according to entanglement, steering  and nonlocality were considered. The hierarchy of quantum correlations in the  increasing order of their strength  was identified as: entanglement, steerability and nonlocality~\cite{gisin1996,wiseman2007,adesso2016}.  For pure states, quantum states are either entangled or separable, for  mixed states, distinctive classification with respect to the aforementioned order of  quantum correlations are more prominent. When correlated quantum states are used as a resource for  teleportation,  teleportation fidelity reveals two different aspects of nonclassicality or measures of correlations of quantum nature. If the teleportation fidelity is greater than $\frac{2}{3}$ (classical limit), the state is non-classically correlated in the sense that it is useful for quantum teleportation. In addition to this, if the fidelity is greater than $F_{lhv}\approx 0.87$~\cite{gisin1996}, then the state is nonlocal in the sense that its teleportation fidelity is incompatible with local hidden variable descriptions, and a state with fidelity greater than $F_{lhv}$   satisfies all the measures of quantum correlations.
In~\cite{Paulson2014,paulson2017}, connection among the different measures of quantum correlations for achieving  quantum teleportation fidelity, and the order of  hierarchy of quantum correlations  were discussed.  \newline

Decoherence of quantum states occurs due to the influence of  noise, and it is known that the order of hierarchy of quantum correlations is preserved~\cite{wiseman2007,Paulson2014,paulson2016} under Markovian noisy channels for different class of pure and mixed states.  The decay of quantum correlations happens in such a way that higher  degree  quantum correlations are lost for a lower value of noise, whereas lower degree QC are lost for higher noise parameters~\cite{paulson2016}. \newline
The classification of quantum channels based on the divisibility properties is quite noteworthy in open system dynamics. According to the divisibility criterion, a quantum channel is Markovian if any intermediate map is completely positive (i.e., if the channel is CP-divisible)~\cite{rivas2014}. In ~\cite{milz2019,shrikant2020}, a broader concept of memory is introduced, whereby CP-divisible quantum processes can occur in non-Markovian regimes as well. CP-divisibility of a quantum process always indicates the lack of information backflow. On the other hand, the absence of P-divisibility can manifest in the form of oscillations in correlation measures such as quantum mutual information, and trace distance, which are monotonic functions of time if the dynamics is P-divisible~\cite{Pradeep}. These oscillations indicate the backflow of information from the environment to the system. Here, we take into account Markovian as well as CP-divisible and P-indivisible non-Markovian channels, and their dynamics are investigated and compared

Quantum speed limit time $(\tau_{QSL})$~\cite{mandelstam1991, margolus1998}, the minimal evolution time between two states, is another quantity that captures Markovianity of the quantum processes. The role of  $\tau_{QSL}$  as a witness of non-Markovianity associated with the non-unitary quantum evolution has been studied \cite{sabrina1,sabrina2}. We investigate the dynamics of $\tau_{QSL}$, and avail its connection with the information backflow to analyse the behavior QC.

%%%%%

%%%
We consider entanglement, quantum steering, and Bell-CHSH nonlocality  as a resource for quantum teleportation \cite{horodecki1996}, and establish their connection with the teleportation fidelity for different class of pure and mixed states in the presence of unital and non-unital noisy channels. This points to the significance of considering the dynamics of two different aspects of nonclassicality/ measures of QC ($F>\frac{2}{3}$ and  $F>F_{lhv}$) associated with the teleportation fidelity along with the entanglement, steering and Bell-nonlocality.  It is known that the  effects of noise on a quantum system are not always detrimental in nature, the revival of quantum correlations occurs due to the backflow of information from the environment to the  system. We show that the decay and  revival of quantum correlations under non-unitary evolution follow the order of hierarchy of QC. Also, we study the quantum speed limit time as a witness of the memory effects of quantum channels. It is shown that dynamics of quantum correlations can be described using $\tau_{QSL}$. Markovian and non-Markovian noisy models of amplitude damping which are non-unital, as well as unital channels such as phase damping, depolarizing and random telegraph noise (RTN) are considered, and noise tolerance of  QC in these cases are discussed.\newline

The work is organized as follows. In  Sec.~\ref{sec1}, we briefly define different measures of quantum  correlations, quantum speed limit time in noisy environment,  and methods to quantify them. In Sec.~\ref{sec2}, the effect of  noisy channels on a quantum system taken to be in a pure entangled state is described, followed by the investigation  of the dynamics of  QC and $\tau_{QSL}$ under the influence of various channels. We establish the connection among different measures of quantum correlations when they are used as a resource for quantum teleportation. A corresponding analysis for initial mixed states is made in Sec. IV. We show that quantum speed limit time can be availed to describe the dynamics of quantum correlations. Results and discussions in Sec.~\ref{sec3} are followed by the concluding section (Sec.~\ref{sec4}).
\section{Quantum correlations}\label{sec1}
Quantum correlations and quantum speed limit time, that can serve  as indicators  of quantumness in a system are defined. Quantum correlations are used as a resource for quantum teleportation. The connections between quantum entanglement, steering and violation of  Bell-CHSH inequality with two different  aspects of nonclassicality associated with the teleportation fidelity are established. In this section, we discuss the methods to estimate different QC for a two-qubit state, $\rho_{AB}$ and the derivation of $\tau_{QSL}$ in open system dynamics.
\subsection{Teleportation fidelity and Bell-CHSH inequality}
In general, a two qubit state is given as,
\begin{equation}
\small
\rho_{AB}=\frac{1}{4}(I_{2}\otimes I_{2}+\sum_{i=1}^{3}r_{i}\sigma_{i}\otimes I_{2}+\sum_{i=1}^{3}s_{i}I_{2}\otimes\sigma_{i} +\sum_{i,j=1}^{3}t_{i,j}(\sigma_{i}\otimes\sigma_{j})).
\label{g_bp_state}
\end{equation}
We have $\sum_{i=1}^{3}r_i=1$ and $\sum_{i=1}^{3}s_i=1$. The correlation matrix is defined as $T=\{t_{i,j}\}$ and the matrix elements $t_{i,j}=Tr[\sigma_{i}\otimes\sigma_{j}\rho]$. 
Two-qubit entangled states are used as a resource for quantum teleportation, and the teleportation fidelity~\cite{horodecki1996} is calculated,
\begin{equation}
F(\rho)=\frac{1}{2}\Bigg(1+\frac{N(\rho)}{3}\Bigg),
\end{equation}
where $N(\rho)=\sum_{i}^{3}u_{i}$; 
$u_{i^{\prime s}}$ are the square root of the eigenvalues of $T^{\dag}T$. The  given state is useful for quantum teleportation 
iff $N(\rho)> 1$, i.e., $F(\rho)>\frac{2}{3}$ (classical limit).\newline
The violation of Bell-CHSH inequality   can be checked by estimating the expectation value of Bell observable $B$~\cite{horodecki1996} for a given state $\rho$, and $B_{\max}=2 \sqrt{\max_{j>k}(u_{j}^{2}+u_{k}^{2})}$. The state $\rho$ violates Bell-CHSH inequality  for $B(\rho)>2$.\newline
\subsection{Quantum Steering}
Quantum steering \cite{einstein1935,wiseman2007steering} makes a reference to the fact that, in the case of biseparable quantum systems, the state of a quantum system can be changed by the action of local measurements on the other system. The degree of steerability of a given quantum state is estimated by considering the amount  by which a steering inequality is maximally violated~\cite{costa2016}. The formula for two qubit-steering is,
\begin{equation}
    S_{n}(\rho)=\max\{0,\frac{\Lambda_{n}-1}{\sqrt{n}-1 }\},
\end{equation}
 $\Lambda_{2}=\sqrt{c^2-c_{\min}^2}$ and $\Lambda_{3}=c$ are steering values in which measurements, $n=2,3$ per party are involved, called two measurement and three measurement steering, respectively. Here, $c = \sqrt{\vec{c}^2}$, $c_{i^{\prime s}}$ are the eigenvalues of correlation matrix $T=\{t_{i,j}\}$ (Eq.~\ref{g_bp_state}), and $c_{\min} \equiv\min {\{|c_i|\}}$.
\subsection{Quantum Entanglement}
We use concurrence~\cite{wootters1998,wootters2001}  as a measure to estimate the entanglement of a quantum state. The concurrence of a  state $\rho$ is defined,
\begin{equation}
C(\rho)=max\{0,\sqrt{\lambda_{1}}-\sqrt{\lambda_{2}}-\sqrt{\lambda_{3}}-\sqrt{\lambda_{4}}\},
\end{equation}
where $\lambda_{i^{\prime s}}$ are the eigenvalues of $\rho\tilde{\rho}$ in the descending order and $\tilde{\rho}=\sigma_{y}\otimes\sigma_{y}\rho^{*}\sigma_{y}\otimes\sigma_{y}$, $\rho^{*}$ is the complex conjugate of the state $\rho$. We have $0<C(\rho)\leq 1$ for entangled states and $C=0$ for separable states.
\subsection{Quantum speed limit time ($\tau_{QSL}$)}  
Quantum speed limit time defines a  bound  on the  minimum  time  required for a quantum system to evolve between two states \cite{Plastino,Kupferman,Deffner}.
The bound on the  quantum speed limit time  for open quantum systems \cite{Plenio,Davidowich}, whose evolution is governed by general quantum channels is,
\begin{equation}
    \tau_{QSL}\geq \frac{2\theta^2}{\pi^2}\frac{\sqrt{tr\rho_{0}^{2}}}{\overline{\sum_{\alpha}{\vert\vert K_{\alpha}(t,0)\rho_{0}\dot{K}_{\alpha}^{\dag}(t,0)\vert\vert}}},
\end{equation}
where $\overline{X}=\tau^{-1}\int_{0}^{\tau} X dt$. $\rho_{0}$ is the initial state, $K_{\alpha^{\prime s}}$ are the Kraus operators characterizing the channel responsible for the evolution of the quantum state, $\vert\vert A\vert\vert=\sqrt{tr(A^{\dag}A)}$ is the Hilbert-Schmidt norm of A, and $\theta=\cos^{-1}( Tr[\rho_{0} \rho_{t}]/Tr[\rho_{0}^2]$). In this work, we investigate the dynamics of $\tau_{QSL}$ for various noisy quantum channels, and the relationship between quantum correlations and speed limit time is demonstrated.
\section{Action of noisy channels}\label{sec2}
\begin{figure}[htbp]
    \centering
    \includegraphics[height=65mm,width=1\columnwidth]{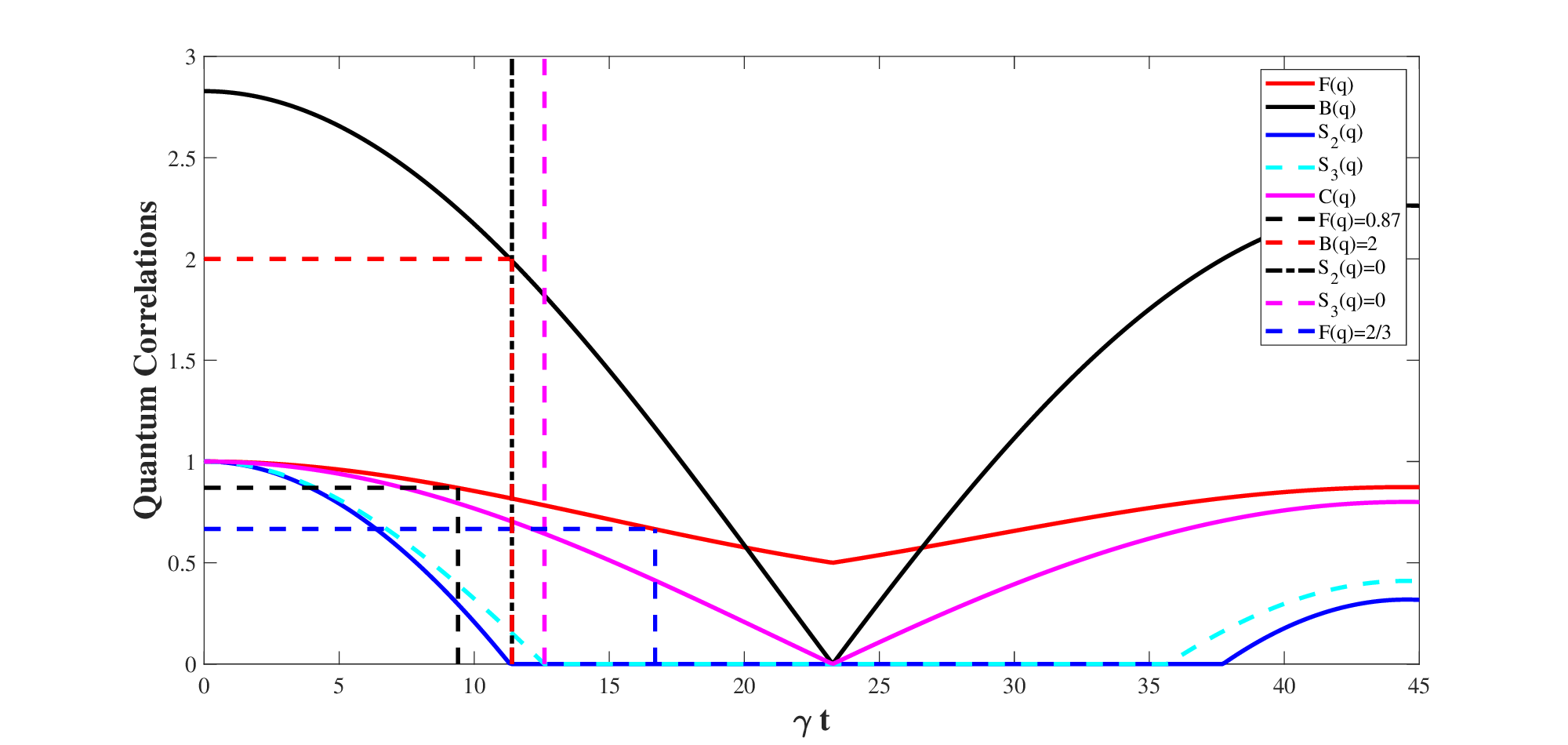}
    \caption{Quantum correlations are plotted as a function of $\gamma t$ $(\Gamma=0.01\gamma)$ in the case of non-Markovian amplitude damping quantum channel acting on maximally entangled bell state.}
    \label{nm_qc_amp}
\end{figure}
The effect of noise on a system can be described using the operator-sum  formalism. We consider various noisy models, both quantum and classical in nature, for example, the amplitude damping channel, phase damping, depolarizing and random telegraph noise (RTN). The evolution of  a quantum system interacting with its environment is,
\begin{equation}
    \rho(t)=\sum_{i}E_{i}(t)\rho(0)E_{i}^{\dag}(t),
\end{equation}
\begin{figure}[htbp]
    \centering
    \includegraphics[height=65mm,width=1\columnwidth]{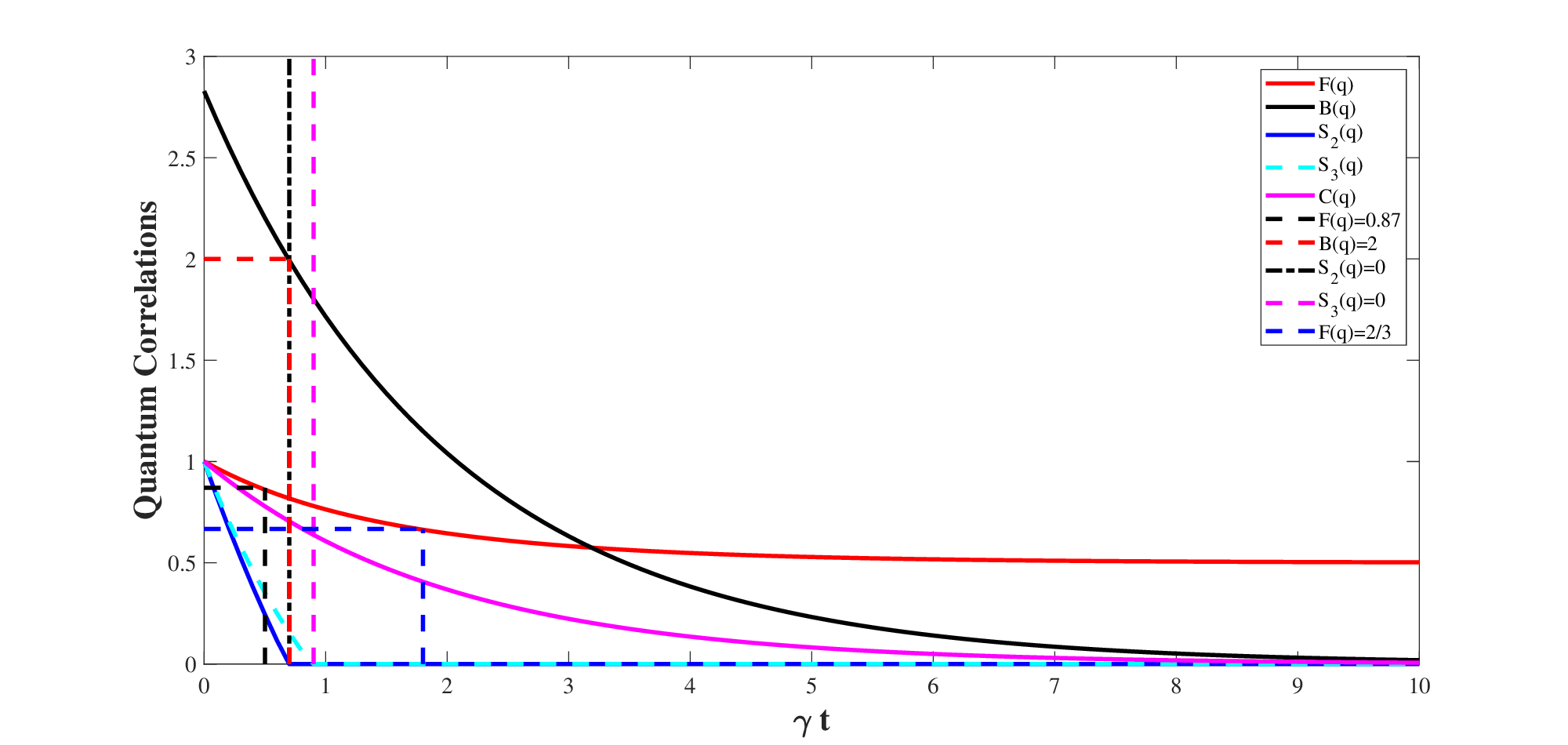}
    \caption{Quantum correlations of Bell state are plotted as a function of $\gamma t$ for Markovian amplitude damping channel.}
    \label{m_qc_amp}
\end{figure}
where, $E_{i^{\prime s}}$ are the Kraus operators characterizing the noise. They satisfy the completeness relation $\sum_{i}E_{i}^{\dag}E_{i}=1$. In general,  local interactions of  a two qubit system with noisy environments can be described as follows,
\begin{equation}
        \rho(t)=\sum_{i,j}E_{i}(t)\otimes E_{j}(t)\rho(0)E_{i}^{\dag}(t)\otimes E_{j}^{\dag}(t).
\end{equation}
Here, we consider the scenario wherein the first qubit interacts with the noisy channel, whereas the second qubit evolves under the noise free condition.
We consider the dynamics of quantum correlations under the influence of different noisy models, both Markovian and non-Markovian (unital as well as non-unital),  and $\tau_{QSL}$ is analyzed for both pure and mixed entangled initial states. A similar dynamics can be observed for the cases where both qubits evolve under noisy quantum channels.
\subsection{Amplitude damping channel}
\begin{figure}[htbp]
    \centering
    \includegraphics[height=65mm,width=1\columnwidth]{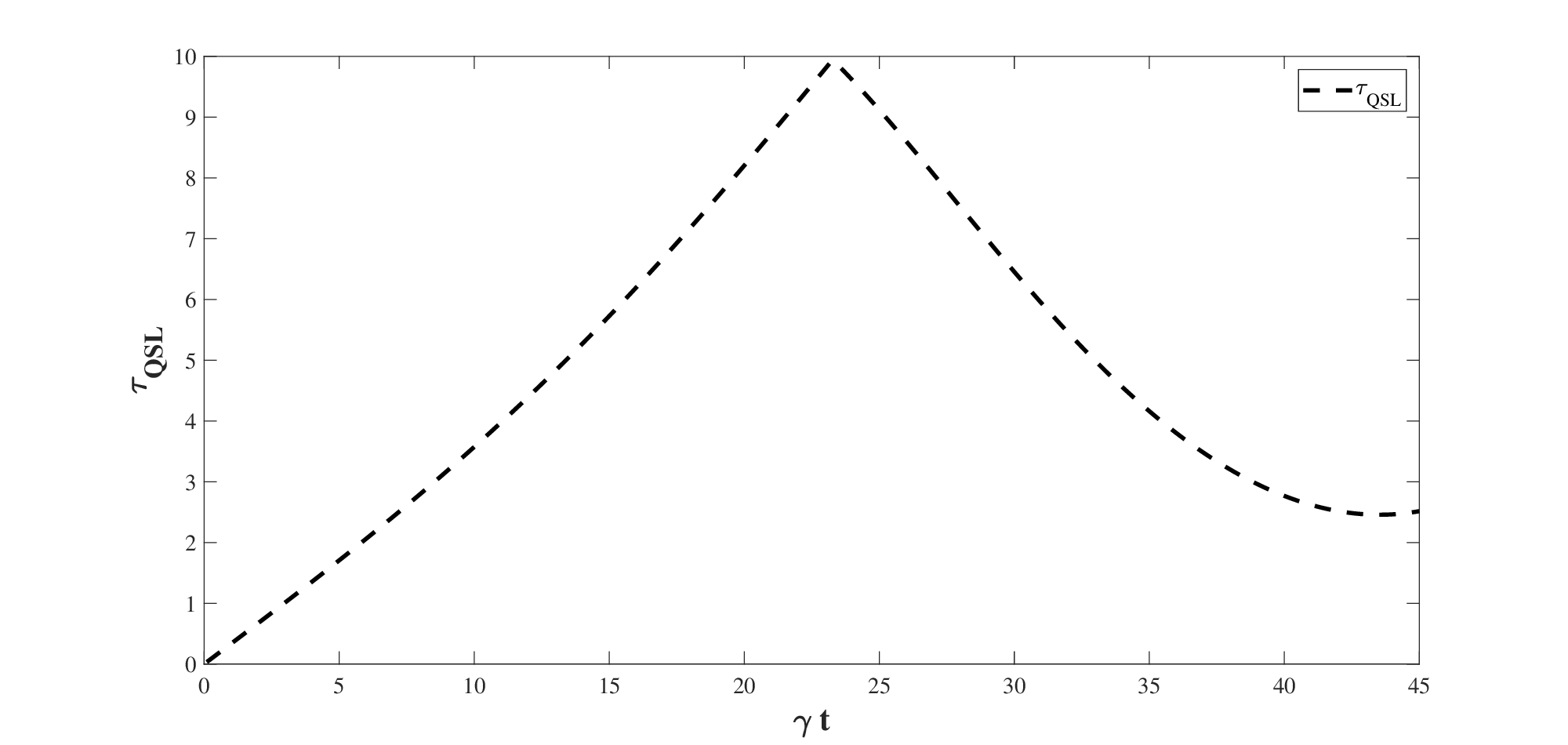}
    \caption{Dynamics of quantum speed limit time of  maximally entangled Bell state for non-Markovian amplitude damping channel ($\Gamma=0.01\gamma$).}
    \label{nm_qsl_amp}
\end{figure}
The Kraus operators of non-Markovian dissipative  quantum channel~\cite{bellomo2007} are given as,
\begin{equation}
    E_{0}=\vert0\rangle\langle0\vert+\sqrt{q}\vert1\rangle\langle1\vert, E_{1}=\sqrt{1-q}\vert0\rangle\langle1\vert,
\end{equation}
we have $q=\exp(-\Gamma t)\{\cos(\frac{dt}{2})+\frac{\Gamma}{d}\sin(\frac{dt}{2})\}^{2}$, $d=\sqrt{2\gamma\Gamma-\Gamma^2}$. Where $\Gamma$ is the line width that  depends on the reservoir correlations time $(\tau_{r}\approx\Gamma^{-1})$ and $\gamma$ is the coupling strength related to qubit relaxation time $\tau_{s}\approx\gamma^{-1}$. The Kraus operators of the amplitude damping channel in the Markovian regime ~\cite{srikanth2008,GPSrik,OmkarSingleQubit} can be obtained by assuming $q=1-\nu$, where $\nu$ is a Markovian exponential decay function .\newline
\begin{figure}[htbp]
    \centering
    \includegraphics[height=65mm,width=1\columnwidth]{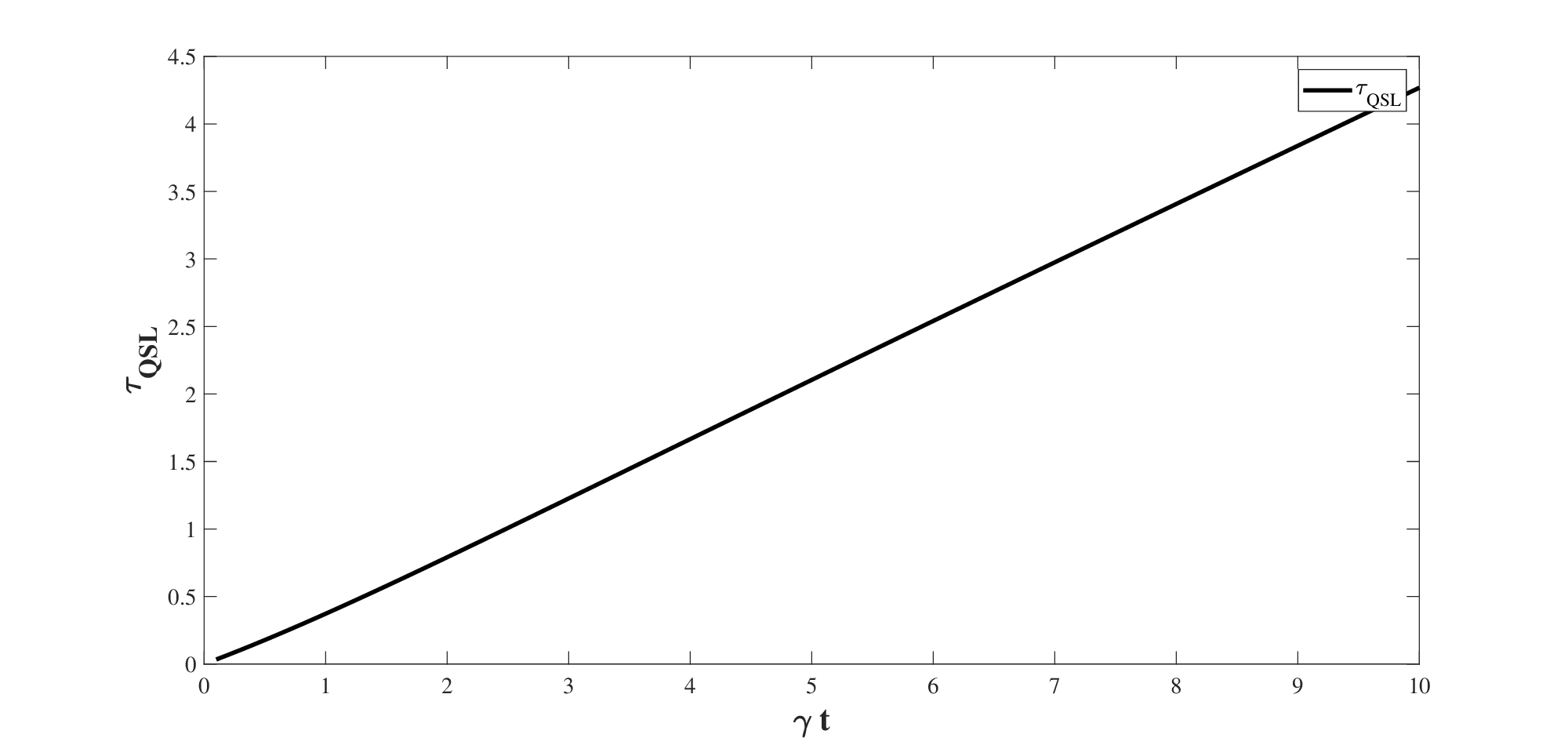}
    \caption{Dynamics of quantum speed limit time of  maximally entangled Bell state for  Markovian amplitude damping channel.}
    \label{m_qsl_amp}
\end{figure}

Let's consider the pure entangled state as  initial state,
\begin{equation}
    \vert\psi\rangle=\alpha\vert 00 \rangle+\beta\vert 11 \rangle,
\end{equation}
where $\vert\alpha\vert^{2}+\vert\beta\vert^{2}=1$. Quantum correlations are calculated, and their dynamics are investigated for maximally entangled Bell state $(\alpha=\beta=\frac{1}{\sqrt{2}})$.

In Fig.~\ref{nm_qc_amp}, the behavior of quantum correlations of maximally entangled Bell state as a function of dimensionless quantity $\gamma t$ under the influence of non-Markovian amplitude damping channel is depicted. We investigate the dynamics of   two different aspects of nonclassicality associated with the quantum teleportation fidelity ( $F(\rho)>\frac{2}{3}$ and $F(\rho)>F_{lhv}(\rho)\approx 0.87$)  along with  entanglement, quantum steering and the violation of Bell-CHSH inequality. QC of  Bell state evolving under the influence of the non-Markovian amplitude damping channel decay initially and revive back after all quantum correlations reach their minimum values, and this process continues.  It is clear from Fig.~\ref{nm_qc_amp} that, the decay and revival of quantum correlations follows a particular order,  higher degree quantum correlations are lost for  small values of channel parameter compared to the  lower degree quantum correlations. The decay of QC, as a function of channel parameter occurs in the following decreasing order, state's teleportation fidelity less than $F_{lhv}\approx0.87$, non-violation of Bell-CHSH inequality, vanishing two and three measures of quantum steering, fidelity less than the classical limit and vanishing entanglement. Thus, $q_{F_{lhv}}\leq q_B \leq q_{S_{2}}\leq q_{S_{3}}\leq q_{T}\leq q_E$, where $q_{F_{lhv}}$, $q_{B}$, $q_{S_{2}}$, $q_{S_{3}}$, $q_{T}$ and $q_{E}$ are the channel parameter values at which, teleportation fidelity becomes less than $0.87$, the states stop violating Bell-CHSH inequality, non-violation of two measure steering inequality, disappearance of three measure quantum steering, teleportation fidelity of states less than the classical limit ($\frac{2}{3}$) and vanishing entanglement (zero concurrence) respectively. Hereafter, we use $q_{F_{lhv}}$, $q_{B}$, $q_{S_{2}}$, $q_{S_{3}}$, $q_{T}$ and $q_{E}$ as the channel parameter values at which corresponding measure of  quantum correlation  fails to capture the quantumness of the state, i.e., $q_{F_{lhv}}$,$q_{B}$, $q_{S_{2}}$, $q_{S_{3}}$, $q_{T}$ and $q_{E}$ are the channel parameter values at which $F(q)=0.87$, $B(q)=2$, $S_{2}(q)=0$, $S_{3}(q)=0$, $F(q)=\frac{2}{3}$ and $C(q)=0$, respectively.   This is considered  for  both the cases of decay and revival of QC interchangeably for all noisy models used in this work. The revival of the quantum correlations occurs in the reverse order, i.e., quantum correlations with lowest degree revives first followed by the restoration of QC with increasing degree of their strength. The revival of quantum correlations follows the order: entanglement, teleportation fidelity greater than the classical limit, steerability of quantum states, violation of Bell-CHSH inequality and teleportation fidelity greater than $F_{lhv}$ ($q_E\leq q_{T}\leq q_{S_{3}}\leq q_{S_{2}}\leq q_B\leq q_{F_{lhv}}$).The dynamics of quantum correlations under the Markovian amplitude channel are shown in Fig.~\ref{m_qc_amp}, sudden death occurs for all quantum correlations except entanglement.  The decay of quantum correlations follows the above discussed order of QC. Here, as expected, revival of QC is not observed. From Figs.~\ref{nm_qc_amp} and~\ref{m_qc_amp}, it is clear that  the decay and  revival of  quantum correlations preserve the hierarchy of non-classical correlations.\newline
\begin{figure}[htbp]
    \centering
    \includegraphics[height=65mm,width=1\columnwidth]{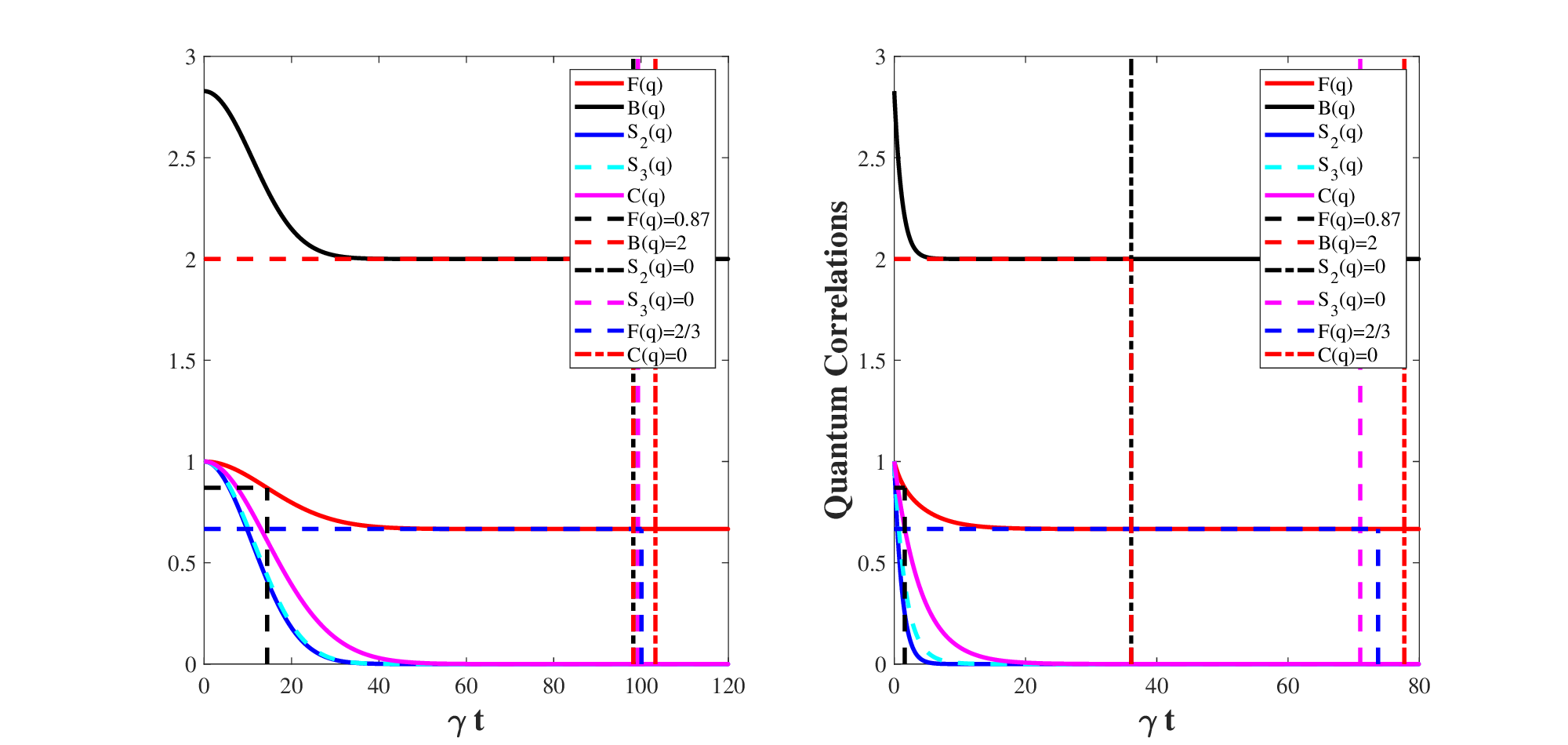}
    \caption{Quantum correlations are plotted as a function of $\gamma t$ for  maximally entangled Bell state in the case of  phase damping channel. \textbf{a)} non-Markovian ($\Gamma=0.01\gamma$)  and \textbf{b)} Markovian.}
    \label{mnm_qc_phd}
\end{figure}
% change the figure for Gamma=0.01 gamma
\begin{figure}[htbp]
    \centering
    \includegraphics[height=70mm,width=1\columnwidth]{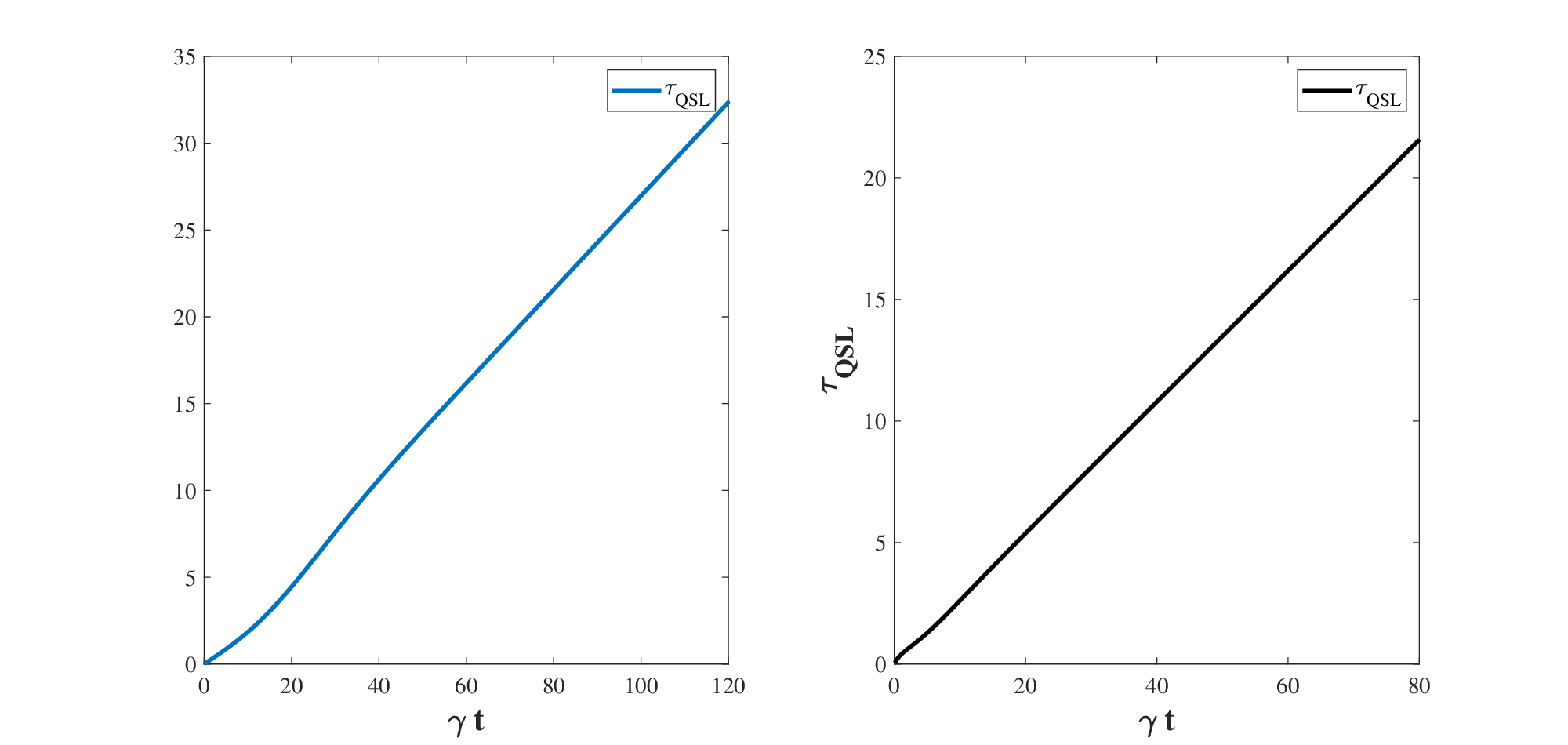}
    \caption{Dynamics of quantum speed limit of maximally entangled Bell states for \textbf{(a)} non- Markovian $(\Gamma=0.01\gamma)$  and \textbf{ (b)} Markovian phase damping channels.}
    \label{mnm_qsl_phd}
\end{figure}
The dynamics of $\tau_{QSL}$ has been considered as a signature of information backflow to the principle system from the reservoir \cite{sabrina1}. From the dynamics of quantum speed limit time (Fig~\ref{nm_qsl_amp}) for non-Markovian amplitude damping channel, it is  clear that $\tau_{QSL}$ increases initially, and starts decreasing after a certain time. The time at which a shift appears in the  dynamics of $\tau_{QSL}$ exactly matches with the time at which revival of lowest degree of quantum correlations starts (Fig.~\ref{nm_qc_amp}). This is interesting,  since the investigation  of $\tau_{QSL}$  reveals the details of non-classical correlations' evolution in the non-Markovian regime.
The connection between  the dynamics of quantum correlations and the speed limit time ascertain the importance of using the latter to analyze the behavior of QC in the case of non-Markovian quantum channels. For Markovian noise, the coupling between the system and reservoir is weak, and hence there is no information backflow and no revival of QC occurs.  This could be inferred from the dynamics of $\tau_{QSL}$. The steady increase of $\tau_{QSL}$ in Fig. \ref{m_qsl_amp} warrants the absence of both information backflow and the revival of quantum correlations for a Markovian approximation.

\subsection{CP-divisible phase damping channel}
We now discuss dephasing quantum channel and its influence on the evolution of quantum correlations. The Kraus operators for a dephasing channel that is historically taken to be non-Markovian ~\cite{yu2010} but is nevertheless P-divisible~\cite{shrikant2020} are:
\begin{equation}
    E_{0}=\vert 0\rangle\langle0\vert+q\vert 1\rangle\langle 1\vert,\, E_{1}=\sqrt{1-q^{2}}\vert 1\rangle\langle 1\vert.
\end{equation}
We have $ q=\exp[\frac{-\gamma}{2}\{t+\frac{1}{\Gamma}(\exp(-\Gamma t)-1)\}]$. $\Gamma^{-1}\approx\tau_{r}$ defines reservoir's finite correlation time and $\gamma$ is the coupling strength related to qubit's relaxation time. In the limit $\Gamma\rightarrow\infty$, phase damping channel reduces to the Markov case, $q=\sqrt{1-\nu}$ identify the Kraus operators for Markovian dephasing quantum channel.\newline 
The behavior of QC of maximally entangled Bell state as a function of $\gamma t$ in the non-Markovian and Markovian  regimes are given in Fig.~\ref{mnm_qc_phd}. In both cases, revival of non-classical correlations does not occur, and the order of decay  satisfies  the same hierarchy as  in the case of  amplitude damping channel, i.e., we have $q_{F_{lhv}}\leq q_{B} \leq q_{S_{2}}\leq q_{S_{3}}\leq q_{T}\leq q_E$. Due to the  memory effects of the non-Markovian phase damping channel,  decay of QC occur slowly as compared to their Markovian counterparts. 
The non-revival of QC for the non-Markovian regime here is due to the noise being CP- divisible and hence also $P$ divisible~\cite{shrikant2020}, which indicates  the absence of backflow. The dynamics of quantum speed limit time for non-Markovian and Markovian phase damping channels is given in~Fig. \ref{mnm_qsl_phd}, and in the absence of revival of QC,  $\tau_{QSL}$ increases steadily in both the cases.

\subsection{Depolarizing quantum channel}
\begin{figure}[htb]
    \centering
    \includegraphics[height=65mm,width=1\columnwidth]{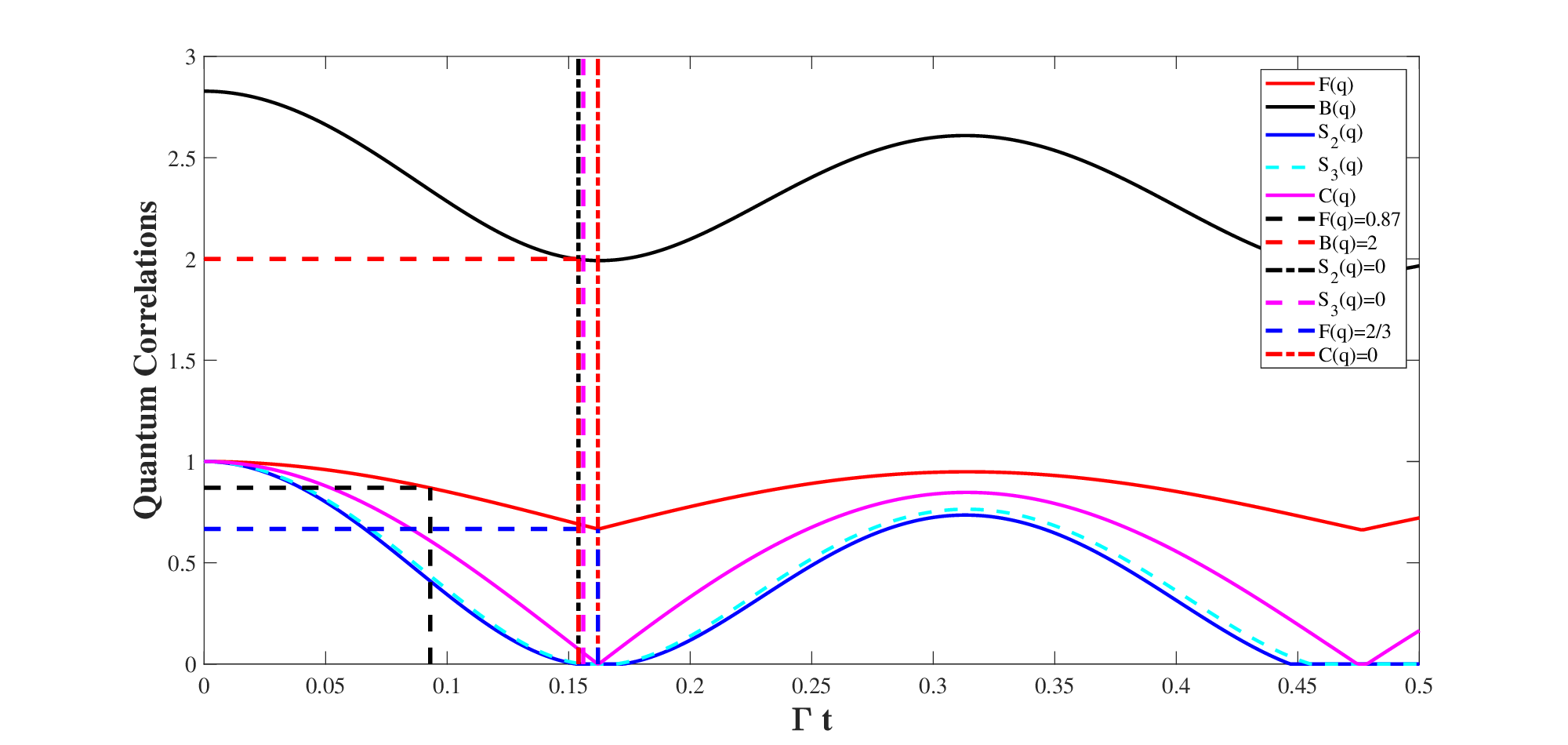}
    \caption{Quantum correlations are plotted as a function of $\Gamma t$ for damped oscillating  non-Markovian depolarizing quantum channel. The values of coupling strengths are chosen as $\gamma_{i}=0.2\Gamma_{i}$,~$i\in\{1,2\}$, $\gamma_{3}=5\Gamma_{3}$ and $\Gamma_{i}=\Gamma$.
  }
    \label{nm_qc_dp}
\end{figure}

\begin{figure}[htb]
    \centering
    \includegraphics[height=65mm,width=1\columnwidth]{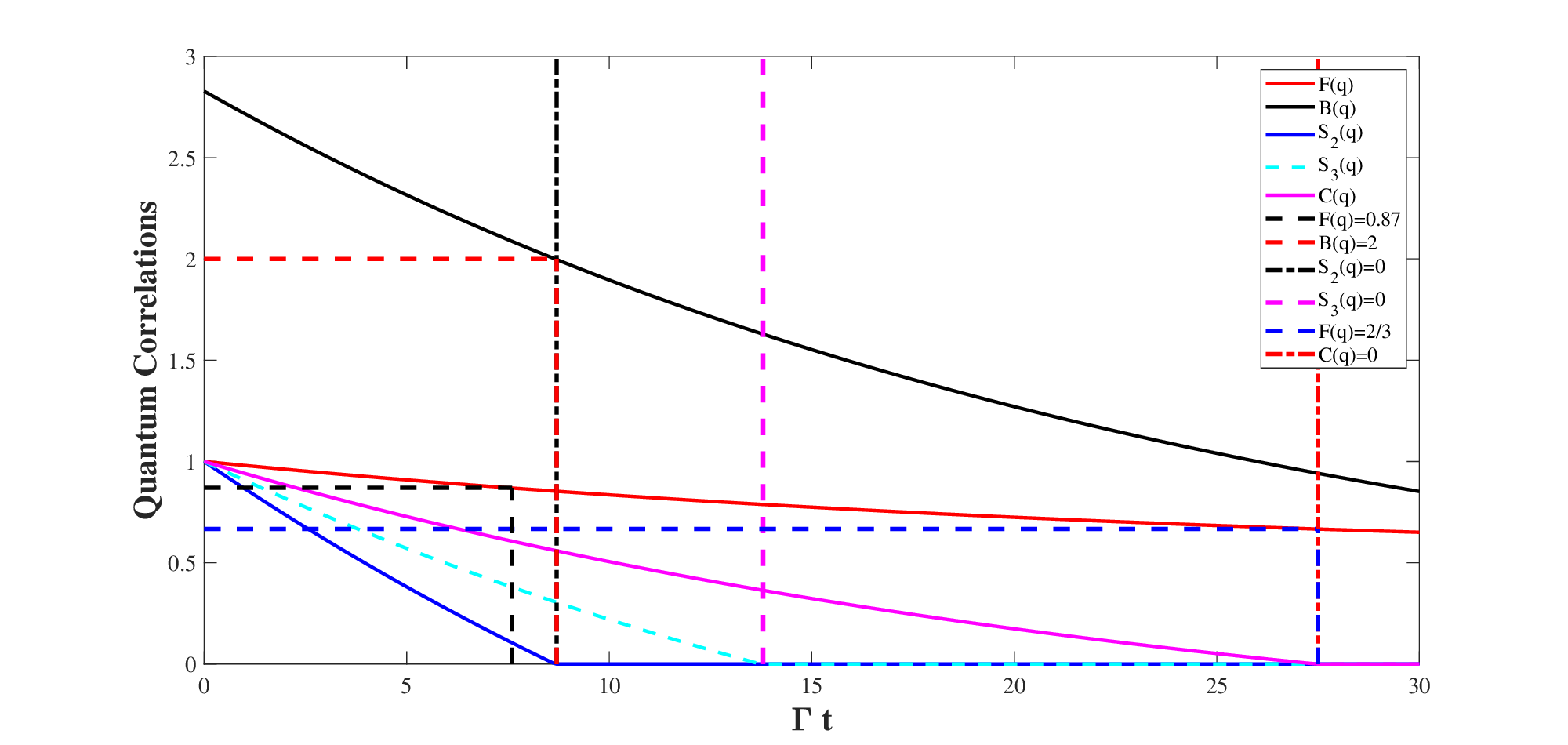}
    \caption{Quantum correlations of Bell state are plotted as a function of $\Gamma t$ for depolarizing quantum channel in the Markovian regime ($\gamma_{i}=0.1\Gamma$). 
  }
    \label{m_qc_dp}
\end{figure}

\begin{figure}[htb]
    \centering
    \includegraphics[height=65mm,width=1\columnwidth]{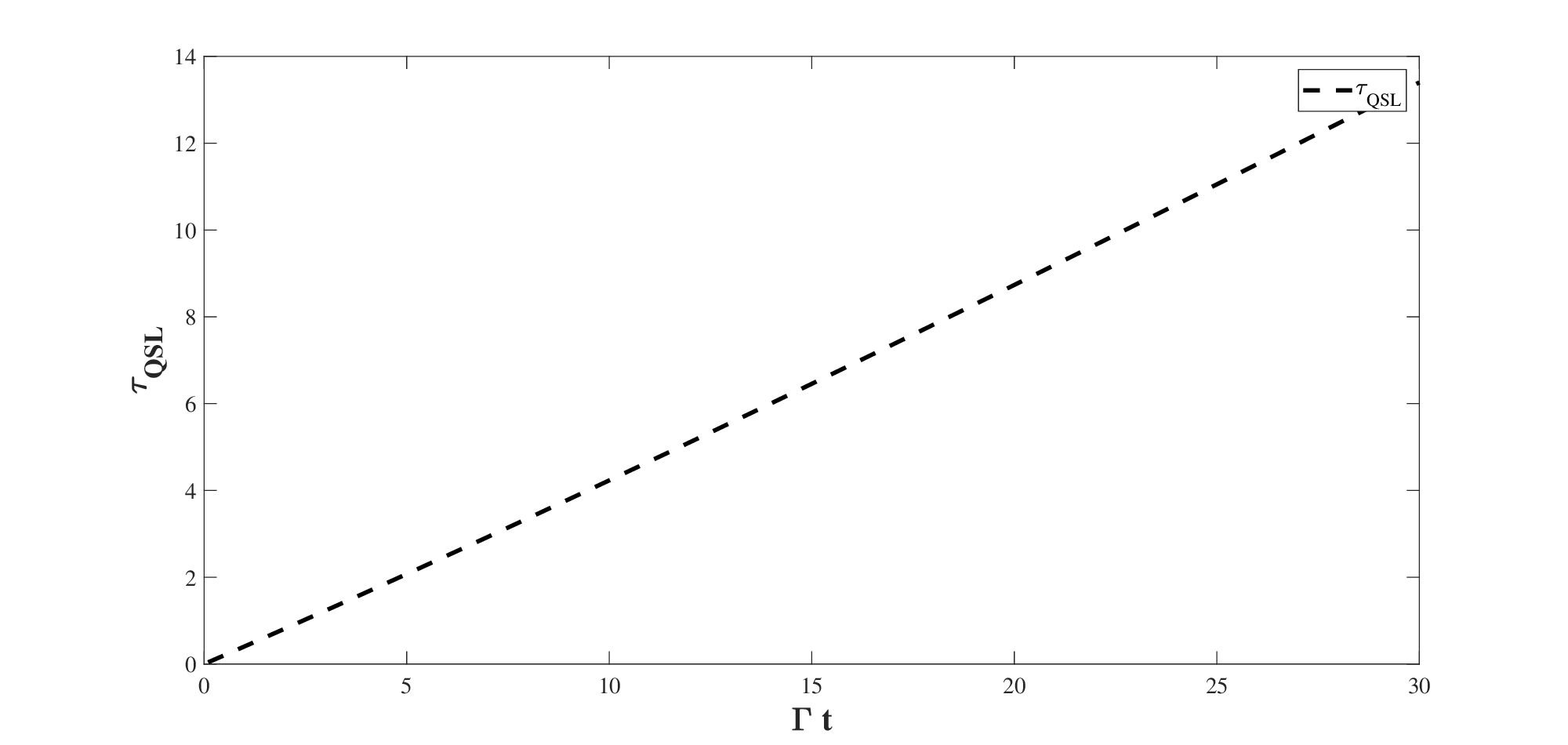}
    \caption{Dynamics of quantum speed limit time of Bell state for depolarizing quantum channel in the Markovian regime ($\gamma_{i}=0.1\Gamma$).}
    \label{nm_m_qsl_dp}
\end{figure}

The Kraus operators of non-Markovian depolarizing quantum channel~\cite{daffer2004} are,
\begin{equation}
    E_{i}=\sqrt{q_{i}}\sigma_{i},
\end{equation}
where $\sigma_{0}=I$, rest of the $\sigma_{i^{\prime s}}$ are the three Pauli's matrices. The complete positivity condition is ensured by identifying the values of $q_{i^{\prime s}}$ as positive, and are given as,
\begin{eqnarray}
\nonumber
q_{0}=\frac{1}{4}[1+\Omega_{1}+\Omega_{2}+\Omega_{3}],\\\nonumber
q_{1}=\frac{1}{4}[1+\Omega_{1}-\Omega_{2}-\Omega_{3}],\\ \nonumber
q_{2}=\frac{1}{4}[1-\Omega_{1}+\Omega_{2}-\Omega_{3}],\\ 
q_{3}=\frac{1}{4}[1-\Omega_{1}-\Omega_{2}+\Omega_{3}].
\label{dppos}
\end{eqnarray}
Here, $\Omega_{i}=\exp(-\frac{\Gamma t}{2})[\cos(\frac{\Gamma d_{i}t}{2})+\frac{1}{d_{i}}\sin(\frac{\Gamma d_{i}t}{2})]$, $d_{i}=\sqrt{(\frac{4 \mu_{i}}{\Gamma_{i}})^2-1}$ with $\mu_{i}^2=\gamma_{j}^2+\gamma_{k}^2$ for $i\neq j\neq k$. Here, $\gamma$ is the coupling strength of the system with the external environment  and $\Gamma^{-1}$ determines the most preferable frequency of the system. The function $\Omega$ has two regimes-pure damping and damped oscillations. $\frac{\mu}{\Gamma}$ determines the behavior of the dynamics. When $0\leq\frac{\mu}{\Gamma}\leq1/4$ the behavior is purely damping. In the regime $\frac{\mu}{\Gamma}>1/4$ damped oscillations exist along with the pure damping. The parameters for which the depolarizing quantum channel is in the Markovian regime are, $\Omega_{i}=e^{-\nu_{i}t}$ and $\nu_{i}=\frac{4}{\Gamma}(\gamma_{j}^{2}+\gamma_{k}^{2})$, here the positivity condition in  Eq.\ref{dppos} is satisfied if and only if $\nu_{i}\leq\nu_{j}+\nu_{k}$.
For an initial Bell state, the evolution of quantum correlations in the non-Markovian regime is depicted in Fig.~\ref{nm_qc_dp}. Differently from non-Markovian phase damping channel, both decay and revival of non-classical correlations happen for non-Markovian depolarizing noise.
\begin{figure}[htb]
    \centering
    \includegraphics[height=65mm,width=1\columnwidth]{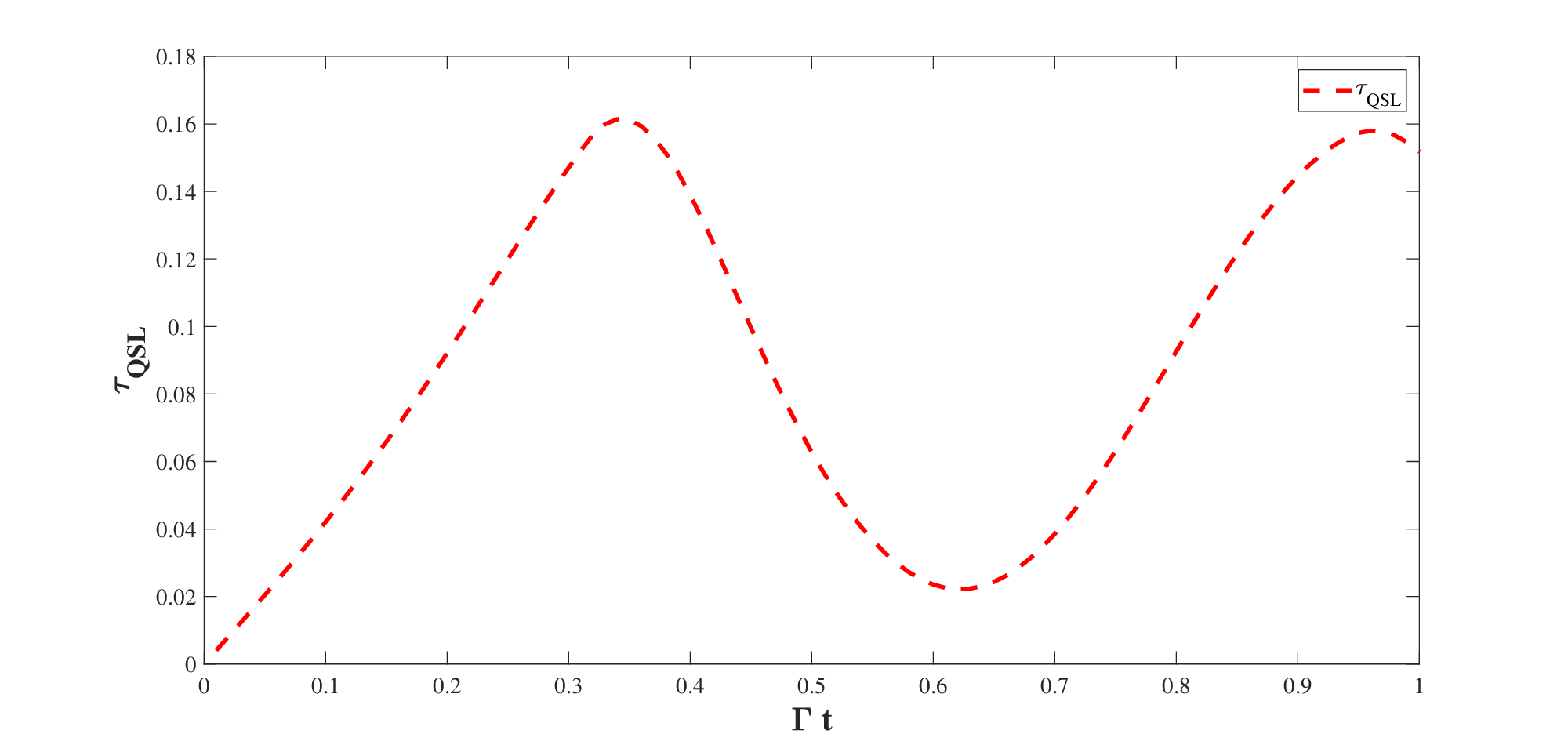} \caption{Dynamics of quantum speed limit time for depolarizing quantum channel in the fluctuating non-Markovian regime. The values of coupling strengths are chosen as $\gamma_{i}=0.2\Gamma_{i}$,~$i\in\{1,2\}$, $\gamma_{3}=5\Gamma_{3}$ and $\Gamma_{i}=\Gamma$.}
    \label{nm_qsl_dp}
\end{figure}
The decay and  revival dynamics of quantum correlations of an entangled initial state under the influence of depolarizing channel is seen to be consistent with the hierarchy of QC.  The channel parameter values for which measures of QC fail to capture the quantumness of the state are in the order $q_{F_{lhv}}\leq q_B \leq q_{S_{2}}\leq q_{S_{3}}\leq q_{T}\leq q_E$. The restoration of quantum correlations occurs in the reverse order ($q_E\leq q_{T}\leq q_{S_{3}}\leq q_{S_{2}}\leq q_B\leq q_{F_{lhv}}$). Initially,  for a small value of noise parameter the state become entangled followed by the creation of other correlations in the increasing order of their strength. Fig.~\ref{nm_qsl_dp} brings out the effect of non-Markovian depolarizing quantum channel on the evolution of quantum speed limit time for Bell state; the oscillatory nature of $\tau_{QSL}$ is the signature of information backflow. In the case of unital non-Markovian depolarizing channel, we do not find  a connection between $\tau_{QSL}$ and QC as seen for the non-unital amplitude damping channel. The decay of QC and the behavior of $\tau_{QSL}$ for depolarizing channel in the Markovian regime are given in Figs.~\ref{m_qc_dp} and ~\ref{nm_m_qsl_dp}, respectively. Purely damping behavior of QC and non-fluctuation of $\tau_{QSL}$ are due to the lack of backflow of information.
\subsection{Random telegraph noise (RTN): P-indivisible phase damping}
\begin{figure}[htb]
    \centering
    \includegraphics[height=65mm,width=1\columnwidth]{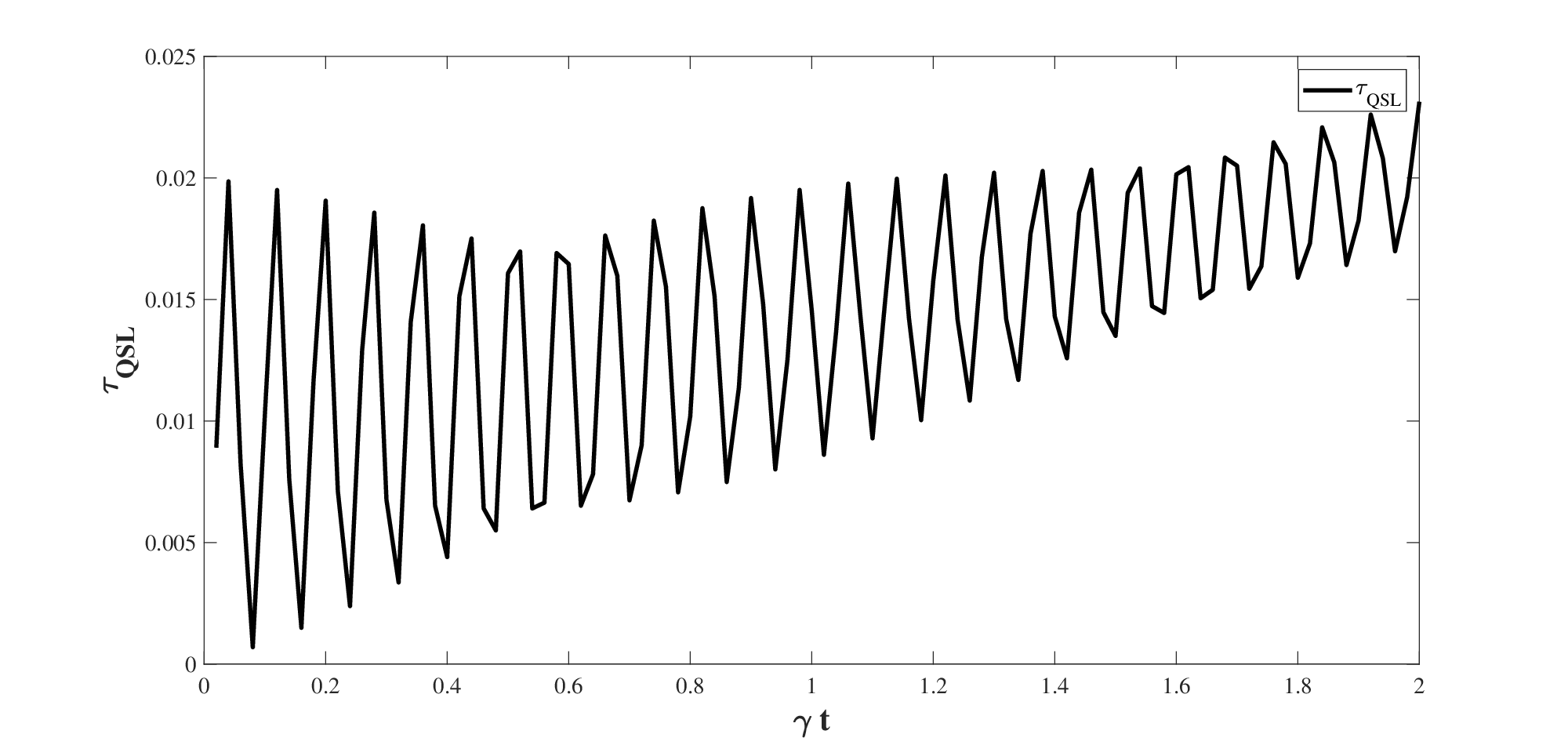}
    \caption{Dynamics of $\tau_{QSL}$ of maximally entangled Bell state  for  non-Markovian random telegraph noise($\frac{a}{\gamma}=40$).}
    \label{nm_qsl_rtn}
\end{figure}

\begin{figure}[htb]
    \centering
    \includegraphics[height=65mm,width=1\columnwidth]{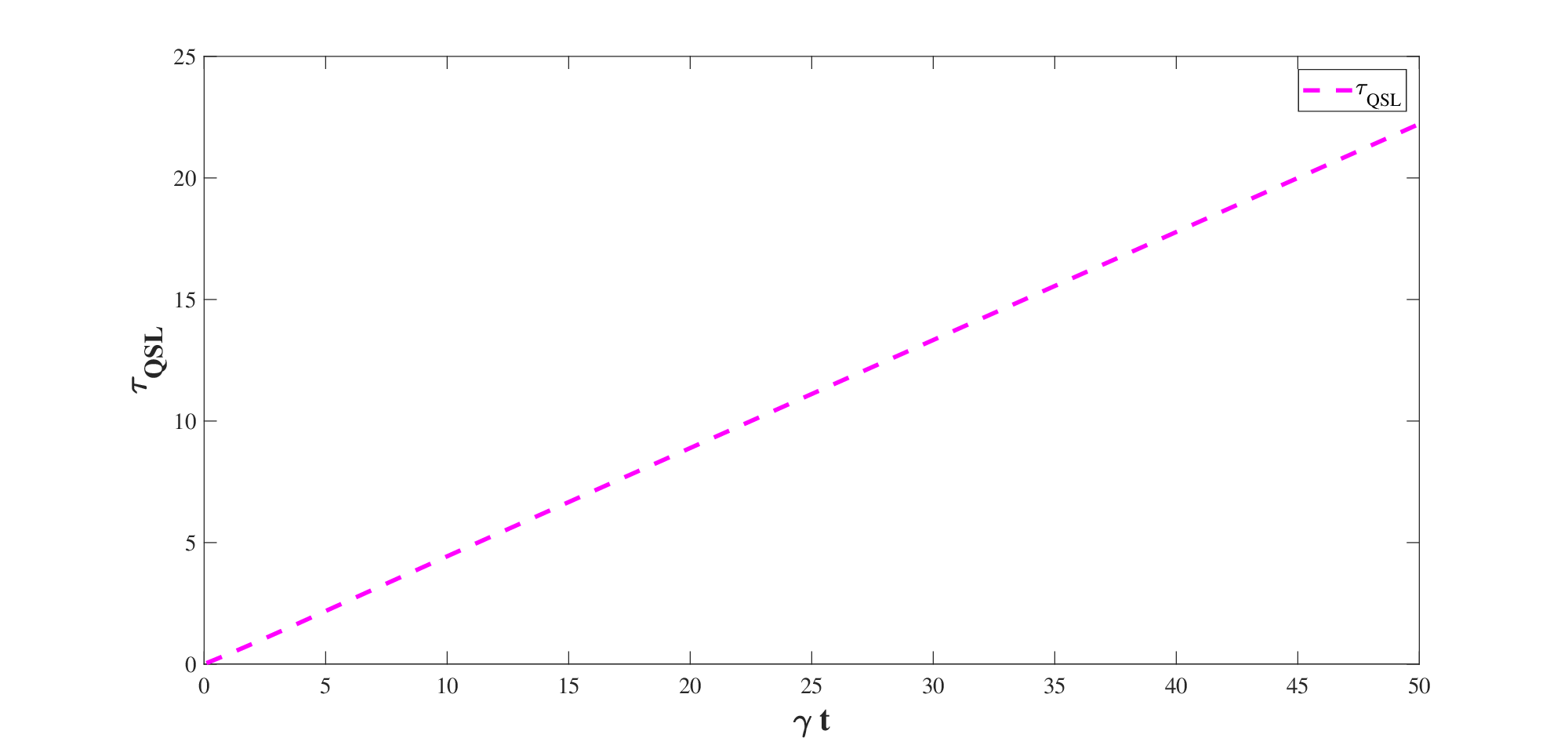}
    \caption{Dynamics of $\tau_{QSL}$ of maximally entangled Bell state  for Markovian random telegraph noise($\frac{a}{\gamma}=0.4$).}
    \label{m_qsl_rtn}
\end{figure}

The quantum dephasing induced by random telegraph noise is now discussed. The Kraus operators representing random telegraph noise~\cite{van1992,pinto2013,mazzola2011,Pradeep}, a P-indivisible phase damping channel are
\begin{eqnarray}
\nonumber
    E_{0}=\sqrt{\frac{1+q(t)}{2}}(|0\rangle\langle0|+|1\rangle\langle1\vert),\\
    E_{1}=\sqrt{\frac{1-q(t)}{2}}(|0\rangle\langle0|-|1\rangle\langle1\vert).
\end{eqnarray}
\begin{figure}[htb]
    \centering
    \includegraphics[height=65mm,width=1\columnwidth]{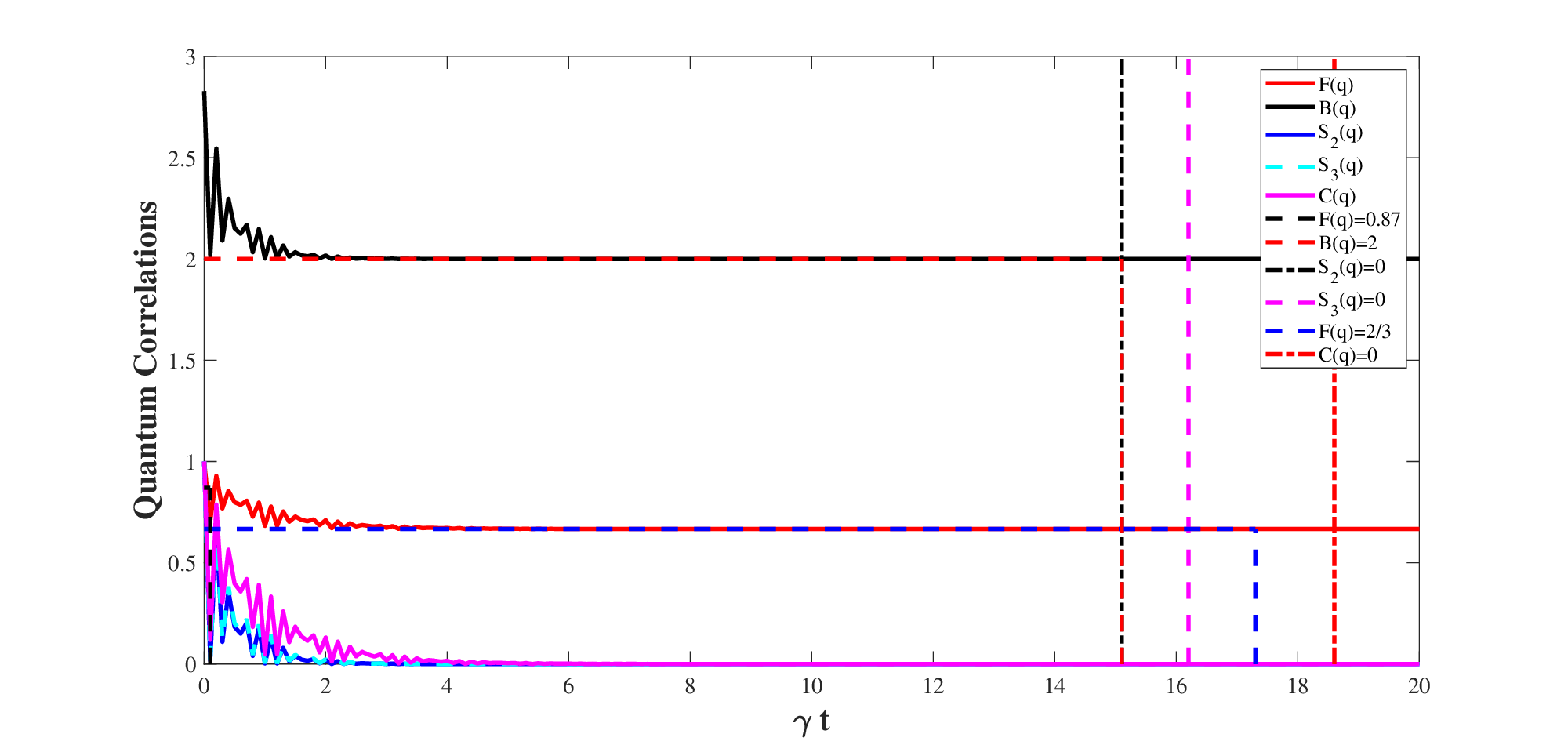}
    \caption{Dynamics of quantum correlations of maximally entangled Bell state  under the influence of non-Markovian random telegraph noise ( $\frac{a}{\gamma}=40$).}
    \label{nm_qc_rtn}
\end{figure}
Where $q(t)$ is the noise parameter based on the damped harmonic oscillator model that accounts the effects of  both Markovian and non-Markovian noise limits on quantum states,

\begin{equation}
    q(t)=e^{-\gamma t}\Big[\cos\Bigg(\sqrt{[(\frac{2a}{\gamma})^2-1]}\gamma t\Bigg)+\frac{\sin\Bigg(\sqrt{[(\frac{2a}{\gamma})^2-1]}\gamma t\Bigg)}{\sqrt{(\frac{2a}{\gamma})^2-1}}\Big].
\end{equation}
\begin{figure}[!htb]
    \centering
    \includegraphics[height=65mm,width=1\columnwidth]{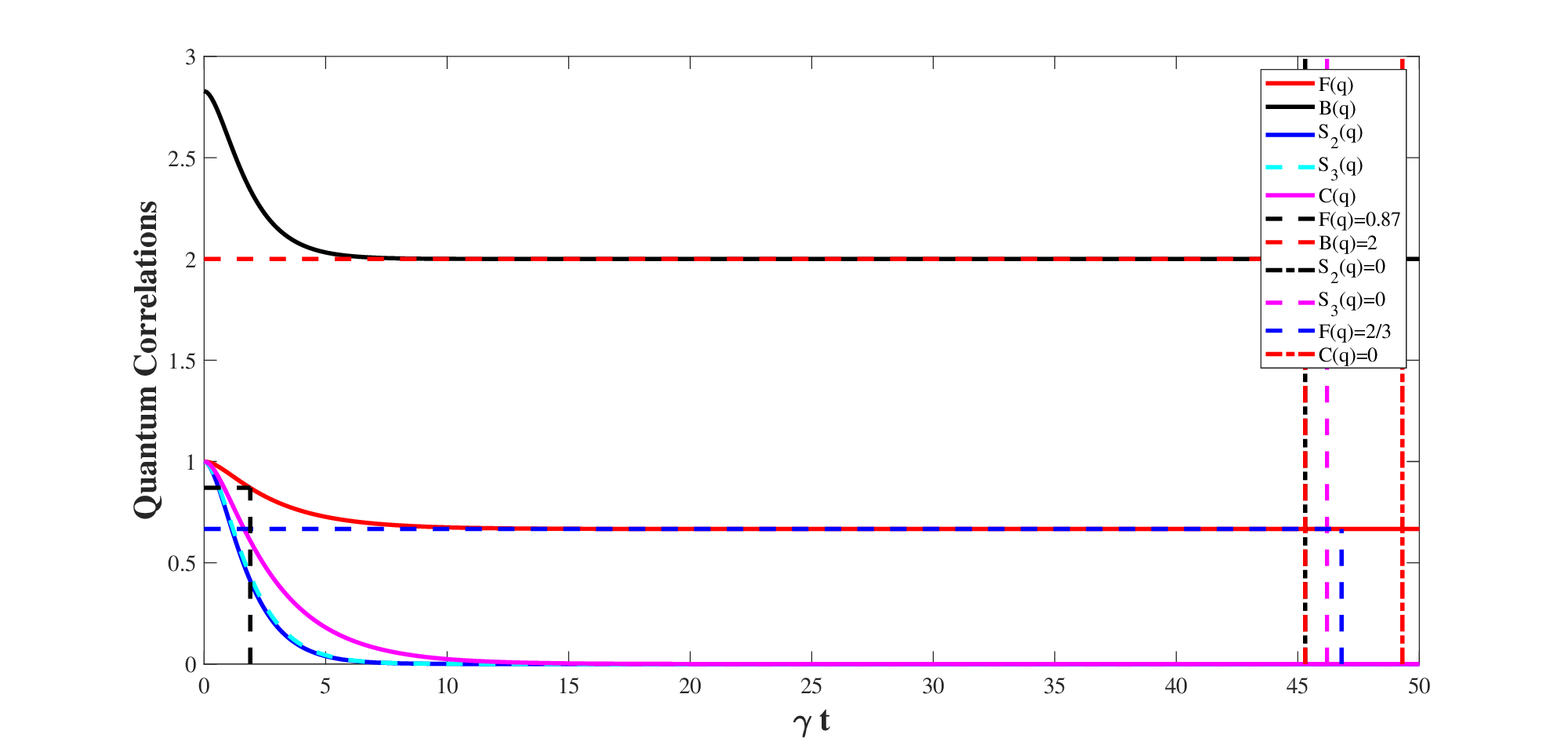}
    \caption{Dynamics of quantum correlations of Bell state  for Markovian random telegraph noise ($\frac{a}{\gamma}=0.4$) .}
    \label{m_qc_rtn}
\end{figure}
The frequency of the harmonic oscillators is $\sqrt{(\frac{2a}{\gamma})^2-1}$. The noise parameter describes two regimes of systems dynamics, for $\frac{a}{\gamma}<0.5$, the channel corresponds to the Markovian dynamics, the purely damping regime  and damped oscillations for $\frac{a}{\gamma}>0.5$ (damped oscillations) corresponds  the non-Markovian evolution. The dynamics of quantum correlations in the non-Markovian regime of RTN channel is shown in Fig.~\ref{nm_qc_rtn}. Initially  all QC fluctuate and decay afterwards. In the Markovian regime (Fig. \ref{m_qc_rtn}), these non-classical correlations decay without fluctuating. The noise parameter values at  which each measure of QC reaches its classical threshold limit obeys the order $q_{F_{lhv}}\leq q_B \leq q_{S_{2}}\leq q_{S_{3}}\leq q_{T}\leq q_E$.
The oscillatory  behavior of  $\tau_{QSL}$ for non-Markovian RTN channel  in  Fig.~\ref{nm_qsl_rtn} captures the presence of  information backflow, whereas quantum speed limit time increases (Fig.~\ref{m_qsl_rtn}) without fluctuation in the Markovian regime.

\section{Mixed entangled states}
\begin{figure}[htb]
    \centering
    \includegraphics[height=65mm,width=1\columnwidth]{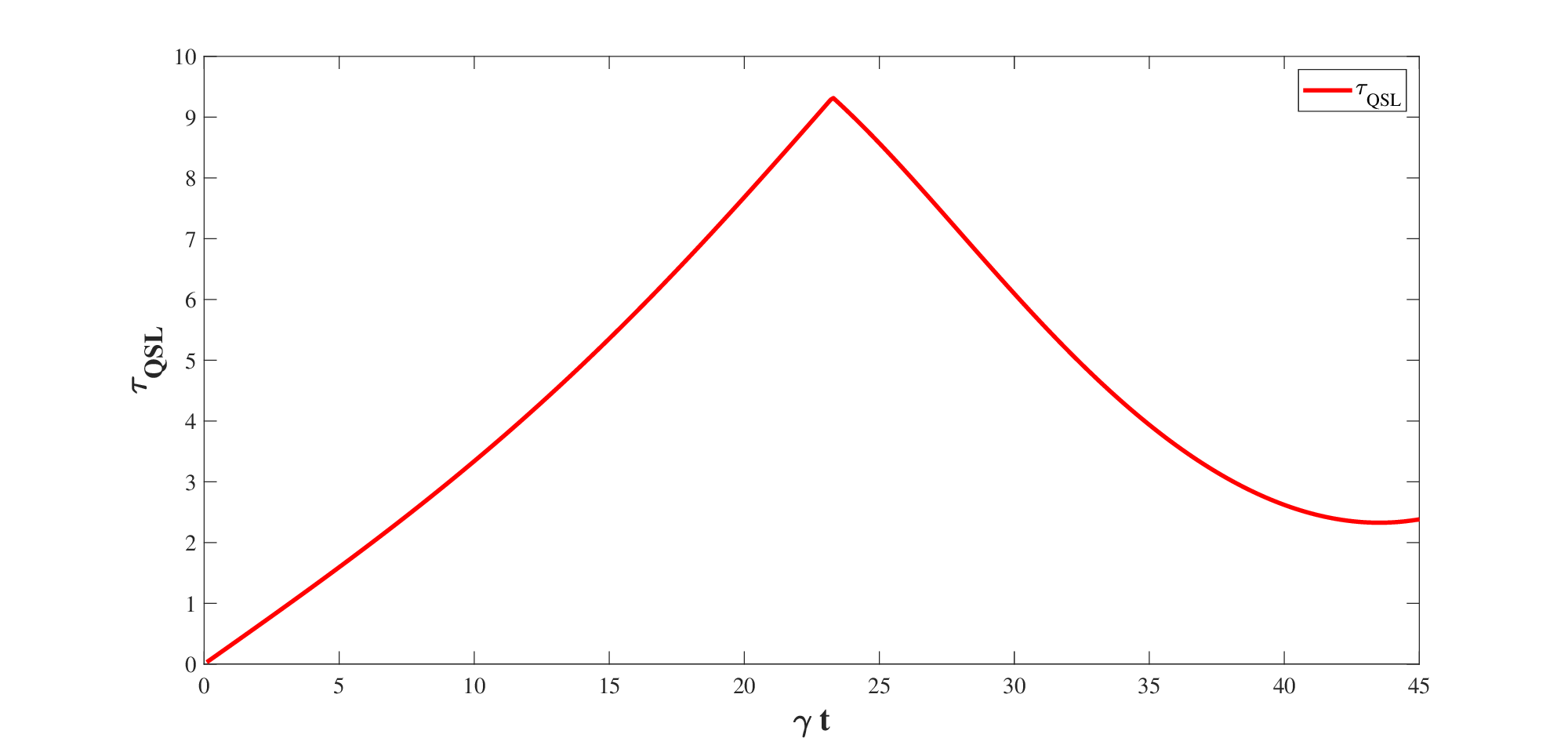}
    \caption{Dynamics of quantum speed limit time of Werner state for $p=0.9$ $(\Gamma=0.01\gamma)$ for non-Markovian amplitude damping channel.}
    \label{nm_qslw_amp_w}
\end{figure}

\begin{figure}[htb]
    \centering
    \includegraphics[height=65mm,width=1\columnwidth]{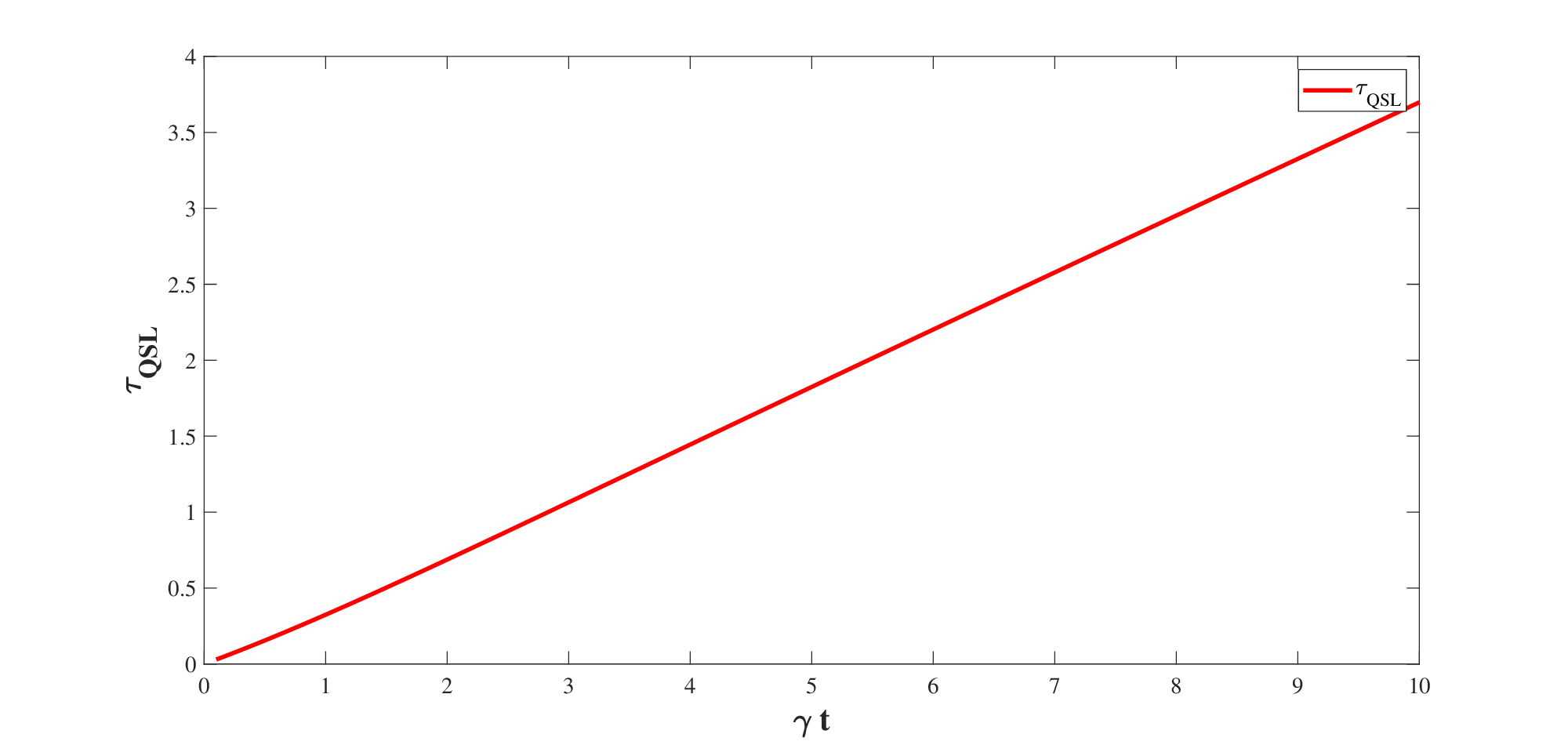}
    \caption{Dynamics of quantum speed limit time of Werner state for $p=0.9$  for Markovian amplitude damping channel.}
    \label{m_qslw_amp_w}
\end{figure}

\begin{figure}[htb]
    \centering
    \includegraphics[height=65mm,width=1\columnwidth]{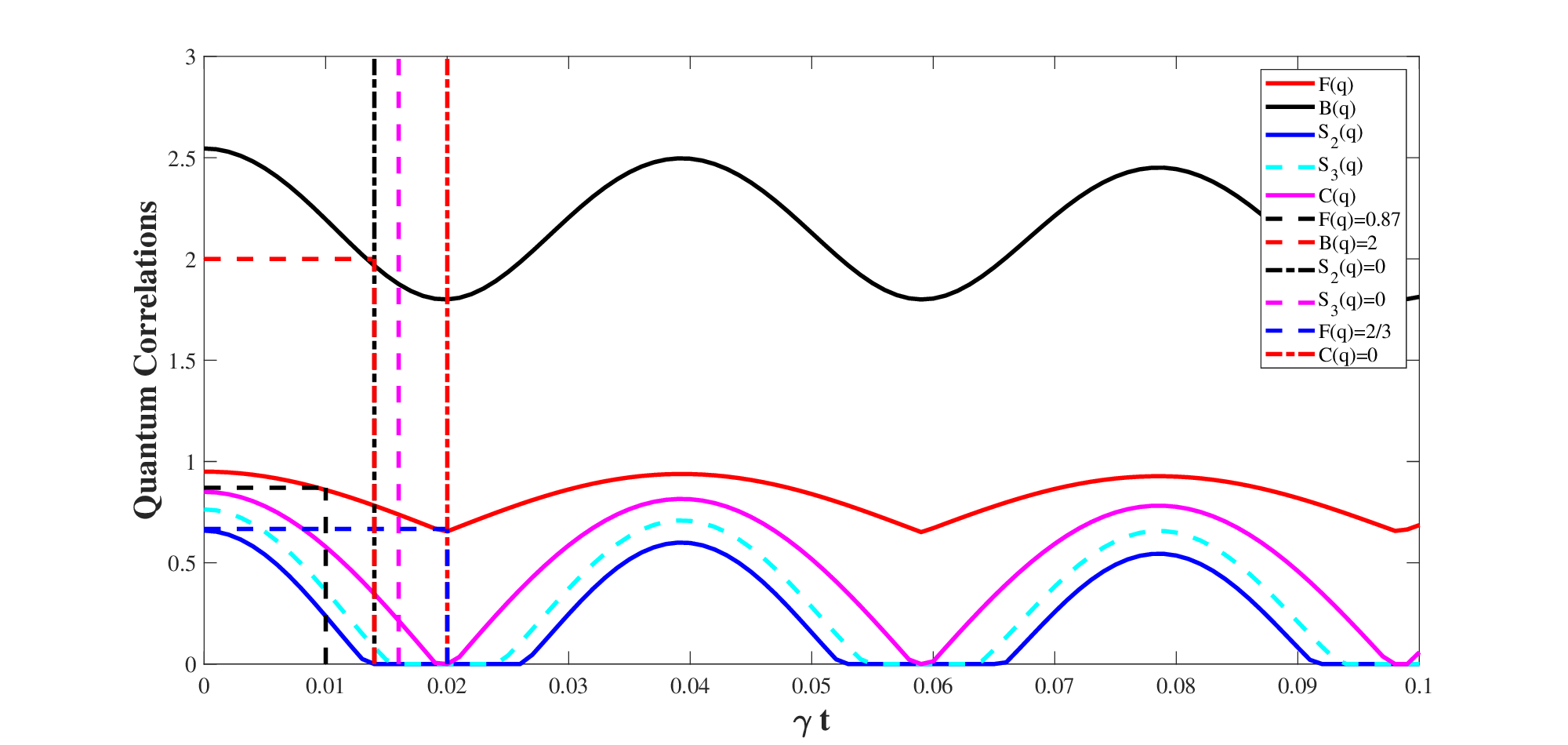}
    \caption{Dynamics of quantum correlations of maximally entangled mixed Werner state for $p=0.9$  for non-Markovian Random Telegraph Noise ($\frac{a}{\gamma}=40$).}
    \label{nm_qc_rtn_w}
\end{figure}

\begin{figure}[htb]
    \centering
    \includegraphics[height=65mm,width=1\columnwidth]{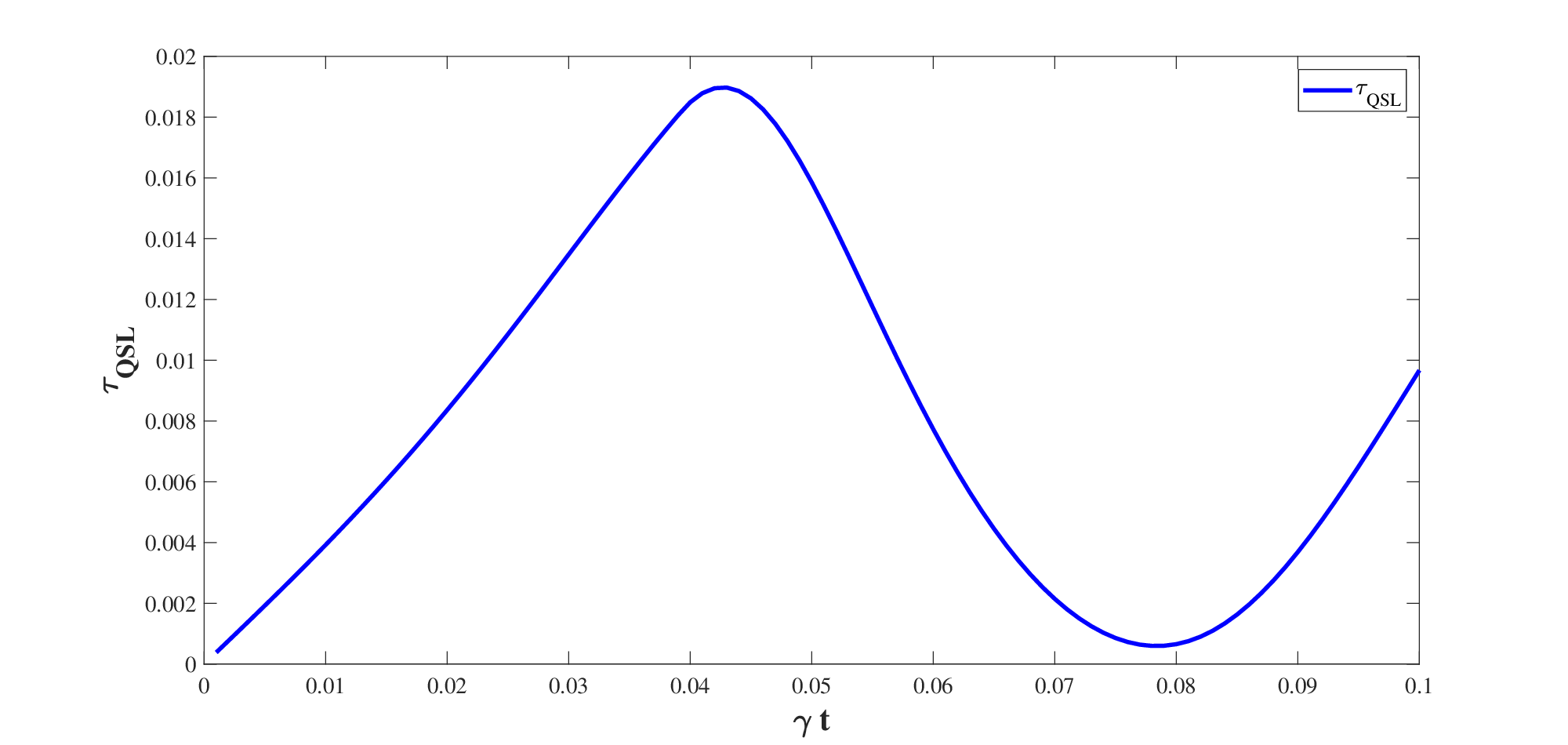}
    \caption{Quantum speed limit time of Werner state for $p=0.9$ as a function of $\gamma t$ for non-Markovian Random Telegraph Noise ($\frac{a}{\gamma}=40$.).}
    \label{nm_qsl_rtn_w}
\end{figure}

The dynamics of quantum correlations and speed limit time for a class of initial mixed states under the influence of different quantum channels are investigated. The initial mixed state we consider  is the Werner state, given as the convex sum of maximally entangled Bell state and maximally mixed separable state
    \begin{equation}
        \rho_{w}=\frac{1-p}{4}I_{4}+p\vert B\rangle\langle B\vert.
    \end{equation}
Here $\vert B\rangle$ can be any one of the four maximally entangled Bell diagonal states. $\rho_{w}$ is entangled for $p>\frac{1}{3}$ and it violates Bell-CHSH inequality and $ST_{2}$ steering for the values $p>\frac{1}{\sqrt{2}}$. Here, we mainly  focus on the  the decoherence effects of amplitude damping and RTN channels on $\rho_{w}$, for a fixed value of mixedness.
\begin{figure}[!htb]
    \centering
    \includegraphics[height=65mm,width=1\columnwidth]{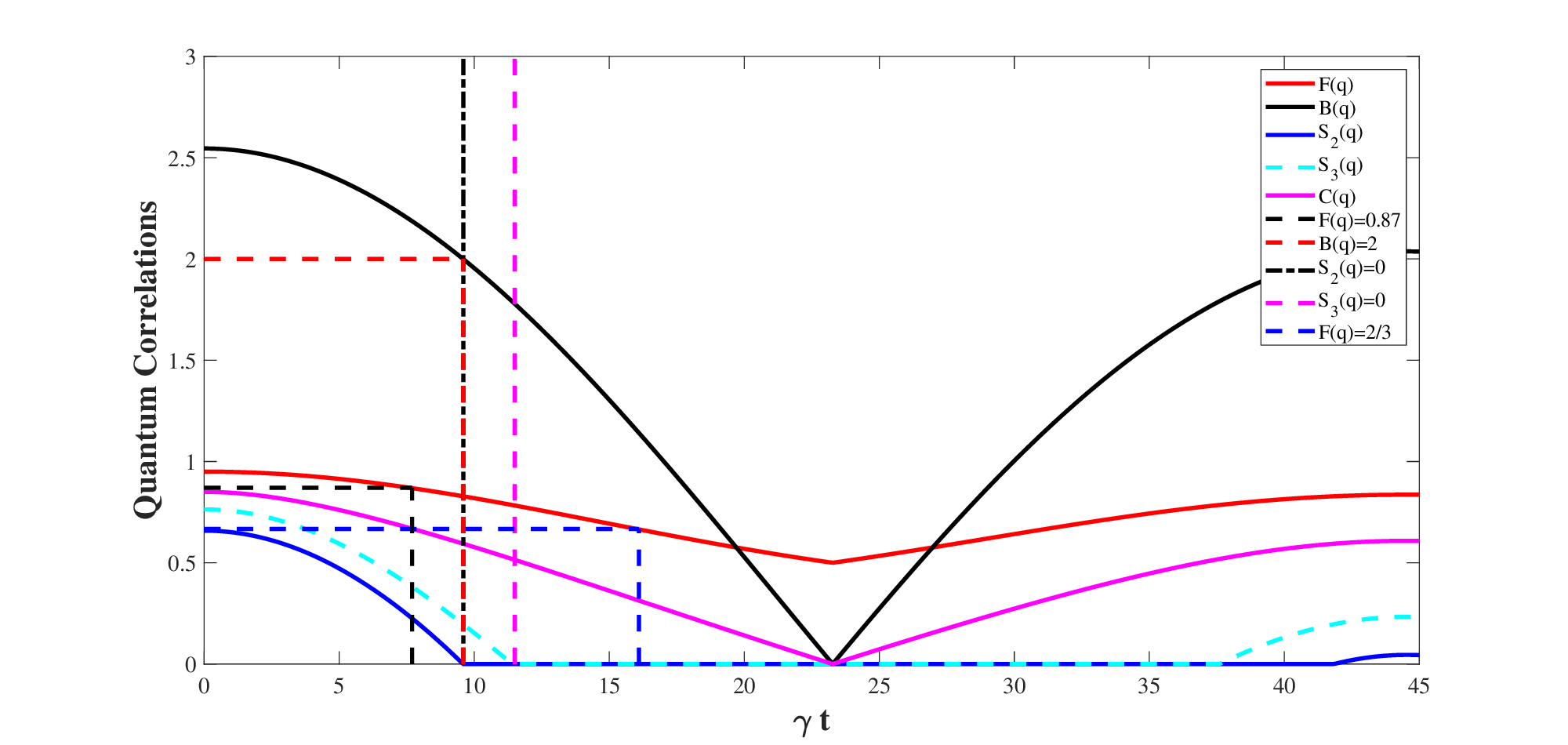}
    \caption{Dynamics of quantum correlations of maximally entangled mixed Werner state for $p=0.9$ $(\Gamma=0.01\gamma)$ under non-Markovian amplitude damping channel.}
    \label{nm_qc_amp_w}
\end{figure}
\begin{figure}[htb]
    \centering
    \includegraphics[height=65mm,width=1\columnwidth]{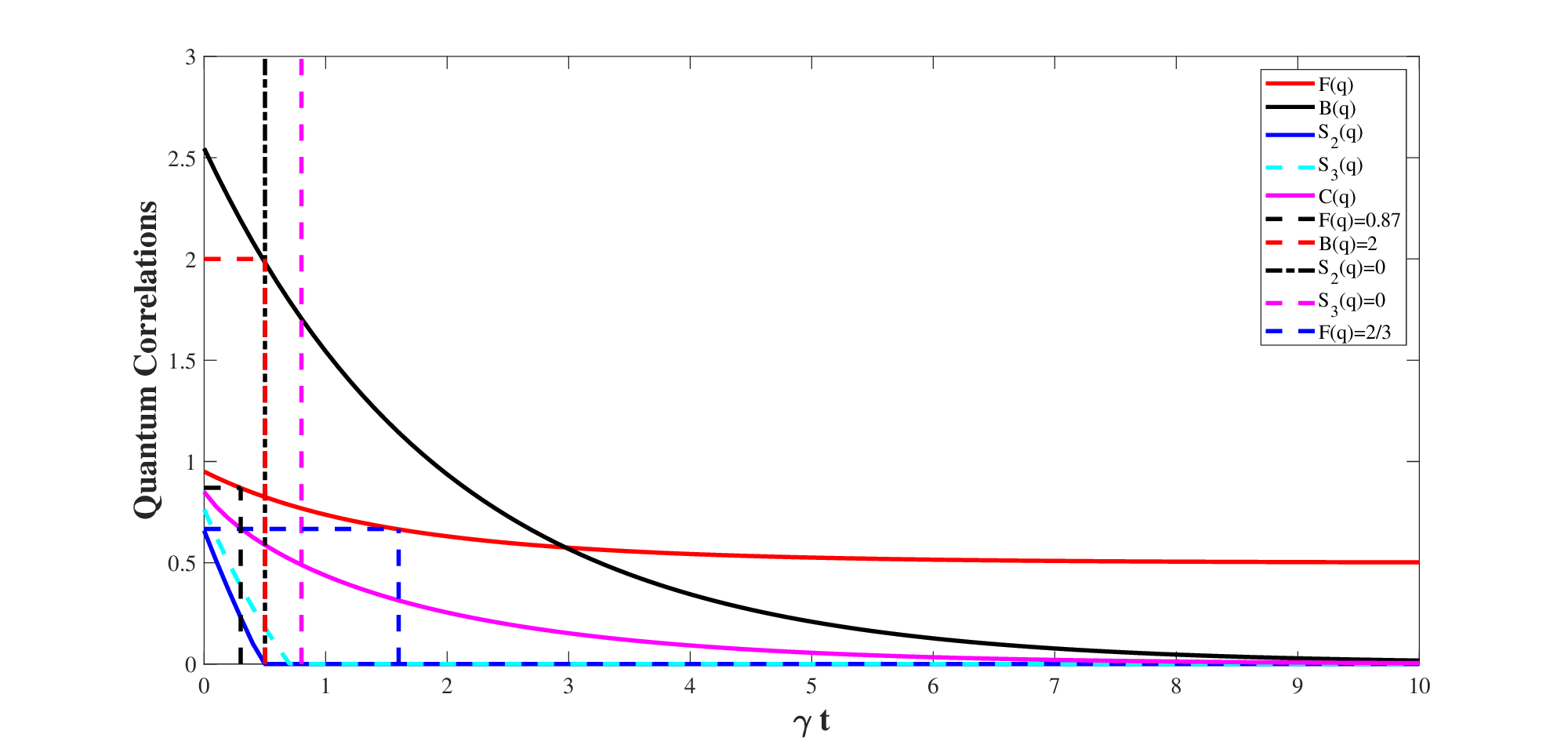}
    \caption{Dynamics of quantum correlations of maximally entangled mixed Werner state for $p=0.9$  under the Markovian  amplitude damping channel.}
    \label{m_qc_amp_w}
\end{figure}
In Figs.~\ref{nm_qc_amp_w} and~\ref{m_qc_amp_w}, the effect of  amplitude damping channel with and without memory on  Werner state for a state parameter value $p=0.9$ is depicted.   As in the case of a pure state                (Fig.~\ref{nm_qc_amp}), in the presence of memory, decay of QC takes place, followed by the revival (Fig. ~\ref{nm_qc_amp_w}). Quantum correlations decay and do not revive (Fig.~\ref{m_qc_amp_w}) in the Markovian regime. The deterioration of QC of mixed state under (non-) Markovian regimes upholds the hierarchy order. Similar to the pure state scenario, $\tau_{QSL}$ (Fig.~\ref{nm_qslw_amp_w}) can be used to analyze the dynamics of non-classical correlations for the non-Markovian  amplitude damping channel. It can be seen that the shift in the nature of $\tau_{QSL}$ matches exactly with the revival of lowest degree quantum correlation (quantum entanglement in the present case) (Fig.~\ref{nm_qc_amp_w}). In the absence of information backflow $\tau_{QSL}$ increases steadily for Markovian noise (Fig.\ref{m_qslw_amp_w}).
QC and $\tau_{QSL}$ of Werner state for non-Markovian RTN channel are depicted in Figs.~\ref{nm_qc_rtn_w} and~\ref{nm_qsl_rtn_w}, respectively. It can be seen from  Fig.~\ref{nm_qc_rtn_w}, that the  strength of QC initially decreases, but due to the system-reservoir coupling and the backflow of information, restoration of non-classical properties of states takes place. The dynamics of non-classical correlation of mixed states under non-Markovian channel is consistent with the order of hierarchy of QC.


\begin{table}
\scalebox{0.9}{
\begin{tabular}{ |c|c|c|c|c| } 
 \hline
 state & noise & Markovian& non-Markovian& decay/revival \\ 
 \hline
\multirow{8}{4em} {Bell state} &\multirow{2}{4em}{AD} & \checkmark & & decay\\ \cline{3-5}
  &  & &\checkmark& both \\ \cline{2-5}
 &\multirow{2}{4em}{PD}  &\checkmark & & decay\\\cline{3-5}
  & & & \checkmark & decay\\ \cline{2-5}
    & \multirow{2}{4em}{DP}  &\checkmark & & decay \\ \cline{3-5}
        & & &\checkmark &  both \\ \cline{2-5}
& \multirow{2}{4em}{RTN}  &\checkmark & & decay \\ \cline{3-5}
        & & &\checkmark & both \\ \hline
 \multirow{4}{4em} {Werner state (p=0.9)} &\multirow{2}{4em}{AD} & \checkmark & & decay\\\cline{3-5}
  &  & &\checkmark& both \\ \cline{2-5}
& \multirow{2}{4em}{RTN}  &\checkmark & & decay \\ \cline{3-5}
      & & &\checkmark & both \\ \hline
\end{tabular}}
\caption{The evolution of quantum correlations under the influence of various unital and non-unital Markovian and non-Markovian quantum channels.}
\end{table}

\section{Results and Discussions}
\label{sec3}
In this paper, we systematically investigated the  dynamics of quantum correlations of two qubit states  that are used as a resource for quantum teleportation in  a noisy environment. We established the connection between quantum correlations and two different aspects of non-classicality associated with the teleportation fidelity. The dynamics of quantum speed limit time can be availed to demonstrate the decay and revival of quantum correlations in the case of memory and memory less quantum channels.  We considered the case of QC in  Markovian as well as CP-divisible and P-indivisible non-Markovian regimes.  From the study of Markovian and non-Markovian channels, it can be inferred that the longevity of quantum correlations gets enhanced due to the memory effects of system-reservoir interaction. The dynamics of QC under the effect of various channels is tabulated in  Table I.
The revival of quantum correlations occurs for all considered non-Markovian channels in the case of both pure and mixed states, except for CP-divisible channels. The non-revival of QC in CP-divisible channels is due to the absence of backflow of information.
In the case of non-unital non-Markovian amplitude damping channel, $\tau_{QSL}$ exactly describes the decay and  revival of quantum correlations. This is not true for unital non-Markovian quantum channels, and is consistent with \cite{sabrina2}. For unital P-indivisible non-Markovian channels, fluctuating $\tau_{QSL}$ and QC imply the existence of information backflow, whereas this is absent in  CP-divisible non-Markovian regime. For a  given Markovian quantum channel, quantum speed limit time increases as time increases, i.e., there occurs no oscillation of $\tau_{QSL}$. This brings forth the marked differences in the behavior of $\tau_{QSL}$ for Markovian, CP-divisible and P-indivisible non-Markovian dynamics \cite{sabrina2}. This is highlighted by the shift in $\tau_{QSL}$ coinciding with the revival of entanglement, for non-Markovian evolution that are $P$ indivisible, exemplified by the non-Markovian amplitude damping channel.
\section{Conclusions}
\label{sec4}
%% write about noise tolerence.
We investigated the effects of  reservoir memory  on the dynamics of quantum correlations of two qubit quantum states. We considered quantum teleportation fidelity, Bell-CHSH function, quantum steering and entanglement as various measures that capture the non-classical aspects of quantum states.  We discussed how these measures of QC are connected with each other under the influence of memory of quantum channels.  We showed the existence of an order of hierarchy in the decay and revival of quantum correlations  under  both Markovian and non-Markovian noises, which is consistent with the previous works. The channel parameter values at which decay of non-classical correlations occur follows the order $q_{F_{lhv}}\leq q_B \leq q_{S_{2}}\leq q_{S_{3}}\leq q_{T}\leq q_E$, whereas the revival of quantum correlations occurs in the reverse order   ($q_E\leq q_{T}\leq q_{S_{3}}\leq q_{S_{2}}\leq q_B\leq q_{F_{lhv}}$) i.e., QC with lowest degree of strength revives first, followed by the revival of correlations in the increasing order of strength. QC revives under all non-Markovian noisy models (P-indivisible)  considered except for the CP-divisible channels, which could be ascribed to the lack of backflow. Noise tolerance of QC under non-Markovian noise is seen to be high compared to that of their Markovian counterpart. We estimated the quantum speed limit time  of states under different  noises and showed that the study of $\tau_{QSL}$ can be used to explain  the characteristic dynamics of QC. Dynamics of QC and $\tau_{QSL}$ were examined for both pure and mixed states in Markovian and non-Markovian regimes. Under Markovian noise, there exists no information backflow and this can be witnessed from the dynamics of quantum speed limit time,  as it increases steadily  as time increases without fluctuations.  Among the non-Markovian noisy models studied here, except for the CP-divisible phase damping noise, fluctuation of $\tau_{QSL}$ was  observed for P-indivisible non-Markovian amplitude damping, depolarizing and RTN channels, which could be attributed to the (non-)existence of information backflow.  It was seen that, for a given  non-Markovian non-unital amplitude damping channel, the dynamics of quantum speed limit time sheds light into the behavior of quantum correlations.  We showed that for non-Markovian amplitude damping noise, the time at which the  lowest degree QC decays exactly matches with the time at which a shift in the behavior of $\tau_{QSL}$  occurs. The  connection between   QC and $\tau_{QSL}$ as seen for non-unital channels cannot be easily established for unital quantum channels and requires further studies.

\section*{Acknowledgement}
SB and RS acknowledge the support from Interdisciplinary Cyber Physical Systems (ICPS) programme of the Department of Science and Technology (DST), India, Grant No.: DST/ICPS/QuST/Theme-1/2019/6.

\bibliography{apssamp}% Produces the bibliography via BibTeX.
%\printbibliography

\end{document}